\documentclass[12pt]{article}
\usepackage{cite}
\usepackage{graphicx}
\usepackage{amssymb}

\def\uu{\langle \bar u u \rangle}
\def\dd{\langle \bar d d \rangle}
\def\sp{\langle \bar s s \rangle}
\def\GG{\langle g_s^2 GG \rangle}
\def\Tr{\mbox{Tr}}
\def\figt#1#2#3{
        \begin{figure}
        $\left. \right.$
        \vspace*{-2cm}
        \begin{center}
        \includegraphics[width=10cm]{#1}
        \end{center}
        \vspace*{-0.2cm}
        \caption{#3}
        \label{#2}
        \end{figure}
	}
	
\def\figb#1#2#3{
        \begin{figure}
        $\left. \right.$
        \vspace*{-1cm}
        \begin{center}
        \includegraphics[width=10cm]{#1}
        \end{center}
        \vspace*{-0.2cm}
        \caption{#3}
        \label{#2}
        \end{figure}
                }

\begin{document}
\title{Meson-Octet Baryon Couplings in the Light
Cone QCD Sum Rules}
\author{T. M. Aliev$^a$ \thanks{Permanent institute: Institute of Physics, Baku, Azerbaijan}
\thanks{taliev@metu.edu.tr},
A. Ozpineci$^a$ \thanks{ozpineci@metu.edu.tr},
~S. B. Yakovlev$^b$,
~V. Zamiralov$^b$ \thanks{zamir@depni.sinp.msu.ru}
\\
{\small
$^a$Middle East Technical University, Ankara, Turkey} \\
{\small
$^b$Institute of Nuclear Physics, M. V. Lomonosov MSU, Moscow, Russia
}}
\begin{titlepage}
\maketitle
\thispagestyle{empty}
\begin{abstract}
The coupling constants of $K$ and $\pi$ mesons with the octet baryons is studied 
in light cone QCD sum rules taking into account  $SU(3)_f$ flavor symmetry breaking
effects, but keeping the $SU(2)$ isospin symmetry intact. It is shown that in the  $SU(3)_f$ flavor 
symmetry breaking case, all the couplings can be 
written in terms of four universal functions instead of $F$ and $D$
couplings which exist in $SU(3)_f$ symmetry case. Comparison of our results
of kaon and pion baryon couplings with existing theoretical and experimental results in the literature
is performed.
\end{abstract}
\end{titlepage}

\section{Introduction}
The analysis of baryon-baryon, baryon-meson scattering and photo-production experiments require the knowledge of the hadronic 
coupling constants involving  mesons. Experimentally, only hadronic coupling constant with pion $g_{NN\pi}$ is 
determined accurately both from nucleon-nucleon and pion-nucleon scattering. However, situation for the kaon case is 
not simple and to reproduce experimental result for kaon-nucleon scattering cross section and kaon photo-production,
many phenomenologically  undetermined coupling constants are needed (see \cite{R1}).
Therefore, it seems a formidable task to determine the kaon-baryon and pion-baryon coupling constants. For this reason, reliable
theoretical approach for estimating these coupling constants is needed. Among all non-perturbative methods, QCD sum 
rules \cite{R2} are especially powerful in studying the properties of hadrons. This method is successfully 
applied to investigation of a variety of problems in hadron physics, in particular the calculation of the meson-baryon
coupling constants. Calculation of the pion-nucleon coupling constants received a lot of attention (see e.g. 
\cite{R3,R4,R5,R6,R7,R8,R9,R10,R11}).

The K meson baryon coupling constants also were studied in the framework of the sum rules (SR) method in \cite{R12,R13,R13p,R14} and in 
light cone SR (LCSR) in \cite{R15}.

The main result of these works is that the predictions of SR for meson baryon couplings depend strongly on the choice of the 
Dirac structure (see e.g. \cite{R9,R10}). Our numerical calculations show that only the $\not\!p \not\!q \gamma_5$
structure leads to a reliable prediction on the meson-baryon coupling constants. 

In this work, we will study pseudoscalar $\pi$ and $K$ meson-baryon 
coupling constants in the framework of an alternative approach to the
traditional SR, i.e. LCSR using the most general expression for hadronic currents as well as $SU(3)_f$ flavor symmetry
breaking strange quark effects. LCSR is based on the operator product expansion near the light cone, which is an expansion
of the time ordered product over the twist rather than the dimension of the operators. Main contribution
in this approach comes from the operators having the lowest twist. The main ingredient of LCSR is the wave
functions of hadrons, which define the matrix element of non-local operators between the vacuum and the one particle 
hadron state (for more see e.g. \cite{R16,R17}).

It should be noted that kaon baryon coupling constants in $SU(3)_f$ symmetry limit have been studied in
\cite{R18} based on the known sum rules for the pion-baryon couplings. 
In the present work we have taken into account the $SU(3)_f$ breaking effects. It is well 
known that in $SU(3)_f$ symmetry limit, the coupling constants of pions and
kaons with baryons are described in terms of two universal constants: $F$
and $D$(see section II). In the absence of this symmetry, it is natural to
ask how much of this structure still remains. One of the central problems
addressed in this paper is the discussion of this question.

The plan of the paper is as follows. In Sect. II, we demonstrate how kaon
baryon coupling constants and the  pion baryon
couplings can be related in the  $SU(3)_f$ symmetry breaking case. 
In Sect. III, 
the LCSR for the meson baryon coupling constants
using the most general form of the baryon currents is derived. Sect. IV is 
devoted to the analysis of the LCSR and 
comparison on our results with the predictions of the other approaches.

\section{Relations Between $K$ and $\pi$ Coupling Constants.}
In this section, we will demonstrate how $K$ and $\pi$ coupling constants to the
baryons can be related.

Let us briefly review the formulas for the coupling constants $KNY$, $K\Xi
Y$ and $\pi \Sigma Y$ ($Y=\Sigma,~\Lambda$) in the $SU(3)_f$ symmetry
limit. In this limit, the interaction Lagrangian can be written as
\begin{eqnarray}
{\cal L} = \sqrt2 \left( D \Tr \bar B \left\{P,B\right\}+ F \Tr \bar B
\left[ P,B \right] \right)
\end{eqnarray}
where
\begin{equation}
B_\beta^\alpha = \left(
\begin{array}{ccc}
	\frac{1}{\sqrt{2}} \Sigma^0 + \frac{1}{\sqrt{6}} \Lambda & \Sigma^+ & p	\\
	\Sigma^- & - \frac{1}{\sqrt{2}}\Sigma^0 + \frac{1}{\sqrt{6}} \Lambda & n \\
	\Xi^- & \Xi^0 & - \frac{2}{\sqrt{6}} \Lambda
\end{array} \right)
\end{equation}
and
\begin{eqnarray}
P_\beta^\alpha = \left(
\begin{array}{ccc}
	\frac{1}{\sqrt2} \pi^0 + \frac{1}{\sqrt{6}}\eta & \pi^+ & K^+ \\
	\pi^- & - \frac{1}{\sqrt2} \pi^0 + \frac{1}{\sqrt6} \eta & K^0 \\
	K^- & \bar K^0  &- \frac{2}{\sqrt6} \eta
\end{array}
\right)
\end{eqnarray}
Here $B_\beta^\alpha$ represent the $\frac12^+$ octet baryons and
$P_\beta^\alpha$ represent the $0^-$ pseudo scalar mesons. From the
Lagrangian, the expressions for the $K$ and $\pi$ couplings can easily be
read off as:
\begin{eqnarray}
g_{\pi^0 p p} &=& F + D,~
g_{\pi^0 \Sigma^+ \Sigma^+} = 2 F,~
g_{\pi^- \Sigma^+ \Sigma^0} = -2 F 
\nonumber \\
g_{\pi^0 \Xi^0 \Xi^0} &=& F - D,~
g_{\pi^+ \Xi^+ \Xi^0} = - \sqrt2 (F-D),~
g_{\pi^{0,\pm} \Sigma^{0,\pm} \Lambda} = \frac{2}{\sqrt3} D
\nonumber \\
g_{K^- p \Lambda} &=& g_{\eta \Xi \Xi} = - \frac{1}{\sqrt3} (3 F + D),~
g_{K^- p \Sigma^0} = D - F
\nonumber \\
g_{K^0 \Xi^0 \Lambda} &=& g_{\eta N N} = \frac{1}{\sqrt3}(3 F - D), ~
g_{K^0 \Xi^0 \Sigma^0} = - (D+F),~ \mbox{etc}
\end{eqnarray}

In order to motivate the treatment in the $SU(3)_f$ violating case, let us
write the $\pi^0$ current as
\begin{eqnarray}
j_{\pi^0} = \sum_{q=u,d,s} g_{\pi^0 q q} \bar q \gamma_5 q
\label{eq5}
\end{eqnarray}
where
$g_{\pi^0 uu} = - g_{\pi^0 dd} = \frac{1}{\sqrt2}$ and $g_{\pi^0 ss} =0$.
Then the coupling of the pion to the $B(qq,q')$ baryon which consists of two identical $q$-quarks and a third different $q'$-quark,
can be written as:
\begin{eqnarray}
\frac{1}{\sqrt2} g_{\pi^0 BB} = g_{\pi^0 qq} 2 F + g_{\pi^0 q'q'} (F-D)
\end{eqnarray}
or explicitly
\begin{eqnarray}
\frac{1}{\sqrt2} g_{pp \pi^0} &=& g_{\pi^0 uu} 2 F + g_{\pi^0 dd}(F-D) = 
\frac{1}{\sqrt2}(F+D)
\nonumber \\
\frac{1}{\sqrt2} g_{\pi^0 \Sigma^+ \Sigma^+} &=& g_{\pi^0 uu} 2 F + g_{\pi^0
ss} (F-D) = \sqrt2 F
\nonumber \\
\frac{1}{\sqrt2} g_{\pi^0 \Xi^0 \Xi^0} &=& g_{\pi^0 ss} 2 F + g_{\pi^0 uu}
(F-D) = \frac{1}{\sqrt2} (F-D)
\end{eqnarray}
etc.

In order to obtain relations between these coupling constants, let us write
formally the coupling constant of the $\pi^0$ to the $\Sigma^0$ as
\begin{eqnarray}
\frac{1}{\sqrt2} g_{\pi^0 \Sigma^0 \Sigma^0} =
g_{\pi^0 uu} F + g_{\pi^0 dd } F + g_{\pi^0 ss} (F-D)
\end{eqnarray}
Note that this coupling is exactly equal to zero. Exchanging, 
first $d \leftrightarrow s$ and then $u \leftrightarrow s$, one obtains two
auxiliary quantities:
\begin{eqnarray}
\frac{1}{\sqrt2}g_{\pi^0 \tilde \Sigma^{0ds} \tilde \Sigma^{0ds}} &=&
g_{\pi^0 uu} F + g_{\pi^0 ss} F + g_{\pi^0 dd} (F-D) = \frac{1}{\sqrt2} D
\nonumber \\
\frac{1}{\sqrt2} g_{\pi^0 \tilde \Sigma^{0 us} \tilde \Sigma^{0us}} &=&
g_{\pi^0 ss} F + g_{\pi^0 dd} F + g_{\pi^0 uu}(F-D) = - \frac{1}{\sqrt2} D
\end{eqnarray}

The following relation holds:
\begin{eqnarray}
g_{\pi^0 \tilde \Sigma^{0ds} \tilde \Sigma^{0ds}} -
g_{\pi^0 \tilde \Sigma^{0 us} \tilde \Sigma^{0us}}  = 2 D =
\sqrt3 g_{\pi^0 \Sigma^0 \Lambda}
\end{eqnarray} 

Up to now, we have considered only the neutral $\pi^0$. To study the
couplings of the charged pions and kaons, let us first define the 
auxiliary ''hyperons'' $\Lambda_{us,ds}$ and $\Sigma^0_{us,ds}$ obtained 
from the normal $\Lambda$ and $\Sigma$ by the exchanges 
$u \leftrightarrow s$ and $d \leftrightarrow s$. Using the wave functions
of $\Lambda$ and $\Sigma^0$ it can be shown that
\begin{eqnarray}
\left\{ \begin{array}{r}
	2 | \Sigma^0_{ds}> = - |\Sigma^0> - \sqrt3 | \Lambda > \\
	2 | \Lambda_{ds}> = - \sqrt3 | \Sigma^0> + | \Lambda >
	\end{array} \right.
\nonumber \\
\left\{ \begin{array}{r}
	2 |\Sigma^0_{us}> = - | \Sigma^0> + \sqrt3 | \Lambda > \\
	2 |\Lambda_{us}> = \sqrt3 | \Sigma^0 > + | \Lambda >
	\end{array} \right.
\end{eqnarray}
where $|\Sigma^0_{ds}>$ and $\Lambda_{ds}$ are the $V=1$ and $V=0$ 
$V$-spin states respectively and $|\Sigma^0_{us}>$ and $|\Lambda_{us}>$
are the $U=1$ and $U=0$ $U$-spin states.

Now, let us write the formal coupling of the $\pi^-$ with $\Sigma^+$ and
the auxiliary $\Lambda_{ds}$
\begin{eqnarray}
2 g_{\pi^- \Sigma^+ \Lambda_{ds}} = - \sqrt3 g_{\pi^- \Sigma^+ \Sigma^0} 
+ g_{\pi^- \Sigma^+ \Lambda} = \frac{2}{\sqrt3}(3 F + D)
\end{eqnarray}
Performing the exchange $d \leftrightarrow s$, $\Lambda_{ds}$ becomes the 
''physical'' $\Lambda$, $\pi^-$ becomes $K^-$ and $\Sigma^+$ becomes
$-p$. Then we get:
\begin{eqnarray}
2 \left( g_{\pi^- \Sigma^+ \Lambda_{ds}} \right)(d \leftrightarrow s)
= - 2 g_{K^- p \Lambda}
= \frac{2}{\sqrt3}(3 F + D)
\end{eqnarray}
which coincide with the $SU(3)_f$ symmetry prediction.

Similarly, writing the coupling of the $\pi^-$ to $\Sigma^+$ and the
auxiliary $\Lambda_{us}$:
\begin{eqnarray}
2 g_{\pi^- \Sigma^+ \Lambda_{us}} = \sqrt3 g_{\pi^- \Sigma^+ \Sigma^0}
+ g_{\pi^- \Sigma^+ \Lambda} = - \frac{2}{\sqrt3} (3 F - D)
\end{eqnarray}
and performing the $u \leftrightarrow s$ exchange, one obtains:
\begin{eqnarray}
2 \left( g_{\pi^- \Sigma^+ \Lambda_{us}} \right)(u \leftrightarrow s) =
2 g_{K^- \Xi^0 \Lambda} = - \frac{2}{\sqrt3}(3 F + D)
\end{eqnarray}

In a similar way, starting from the formal couplings 
$g_{\pi^- \Sigma^+ \Sigma^0_{ds}}$ and $g_{\pi^- \Sigma^0 \Sigma^0_{us}}$,
and performing the $d \leftrightarrow s$ and $u \leftrightarrow s$ exchanges
respectively, one can obtain the following relations:
\begin{eqnarray}
-2 \left( g_{\pi^- \Sigma^+ \Sigma^0_{ds}} \right)(d \leftrightarrow s)
= 2 g_{K^- p \Sigma^0} = 2 (-F + D)
\nonumber \\
-2 \left( g_{\pi^- \Sigma^+ \Sigma^0_{us}} \right)(u \leftrightarrow s)
= 2 g_{K^0 \Xi^0 \Sigma^0} = - 2 (F+D)
\end{eqnarray}

These expression show how we can construct sum rules for the $K$ 
baryon coupling constants, starting from the corresponding sum rules for the
$\pi$ baryon coupling constants. 

\section{Light Cone QCD Sum Rules for the Meson Baryon Couplings}
In this section we will derive light cone sum rules for the meson baryon couplings.  Sum rules for the meson-baryon couplings 
can be obtained by equating two different representations of a suitably chosen correlator, written in terms of hadrons and
quark-gluons. We begin our calculation by constructing the following correlator:
\begin{eqnarray}
\Pi^{B_2 \rightarrow B_1 {\cal M}} = i \int d^4x e^{i p x} \langle {\cal M}(q) \vert {\cal T} \eta_{B_1}(x) {\bar \eta}_{B_2}(0) \vert 0 \rangle
\label{corrf}
\end{eqnarray}
where ${\cal M}$ is either a pion or a kaon, $\eta_B$ is the interpolating current 
of the baryon under consideration, ${\cal T}$ is the time ordering operator, and $q$ is 
the momentum of the ${\cal M}$-meson. This correlator can be calculated on one side phenomenologically, in terms of the hadron 
parameters, and on the other side by the operator product expansion (OPE) in the deep Euclidean region $p^2 \rightarrow -
\infty$, using the quark gluon language. By matching both representations through the dispersion relations, one
obtains the sum rules.

Let us firstly discuss the phenomenological part of the correlator function Eq. (\ref{corrf}).
Saturating the correlator function by ground state baryons with quantum numbers of the corresponding baryons, we get
\begin{eqnarray}
\Pi^{B_2 \rightarrow B_1 {\cal M}} (p_1^2,p_2^2) = 
\frac{ \langle 0 \vert \eta_{B_1} \vert B_1(p_1) \rangle }{p_1^2-M_1^2} 
\langle B_1(p_1) {\cal M}(q) \vert B_2(p_2)\rangle
\frac{ \langle B_2(p_2) \vert \eta_{B_2} \vert 0 \rangle }{p_2^2-M_2^2}
+\cdots
\label{phen}
\end{eqnarray}
where $p_2 = p_1 +q$, $M_i$ is the mass of the baryon $B_i$, and $\cdots$ stand for the
contributions of the higher states and the continuum.

The matrix elements of the interpolating currents between the vacuum and a single baryon state, 
$B_i$, with momentum $p$ and having spin $s$ is defined as:
\begin{eqnarray}
\langle 0 \vert \eta_{B_i} \vert B_i(p,s) \rangle = \lambda_{B_i} u(p,s)
\label{residue}
\end{eqnarray}
where $\lambda_{B_i}$ is the overlap amplitude and $u$ is the Dirac spinor 
for the baryon. In order 
to write down the phenomenological part of the sum rules from Eq. (\ref{phen})
it follows that one also needs the
matrix element $\langle B_1(p_1) {\cal M}(q) \vert B_2(p_2)\rangle$. This matrix 
element is defined as:
\begin{eqnarray}
\langle B_1(p_1) K(q) \vert B_2(p_2)\rangle =
g_{B_2B_1{\cal M}} \bar u(p_1) i \gamma_5 u(p_2)
\label{vertex}
\end{eqnarray}
Using Eqs. (\ref{residue}) and (\ref{vertex}) and summing over the baryons' spins, 
we get the following phenomenological representation of the correlator:
\begin{eqnarray}
\Pi^{B_2 \rightarrow B_1 {\cal M}} (p_1^2,p_2^2) &=& i \frac{g_{B_2 B_1 {\cal M}} \lambda_{B_1} 
\lambda_{B_2}}{(p_1^2 - M_1^2)(p_2^2 - M_2^2)} \left( - \not\!p \not\!q
\gamma_5 - M_1 \not\!q \gamma_5 \right. 
\nonumber \\
&+& \left. (M_2 - M_1) \not\!p \gamma_5 + (M_1 M_2 -
p^2) \gamma_5 \right) + \cdots
\label{phen_rep}
\end{eqnarray}
where $\cdots$ stands for the contribution of the higher states and the 
continuum. From Eq. (\ref{phen_rep}), it is
seen that the correlator has numerous structures and in principle any 
structure can be used for the 
determination of the meson-baryon coupling constant. But our numerical 
analysis show that the sum rules
obtained from $\not\!q \gamma_5$ and $\gamma_5$ do not converge and hence
can not be used for a reliable 
determination of the coupling constant. The sum rules obtained from the 
structure $\not\!p \gamma_5$ do 
converge. But due to the small factor $M_1 - M_2$ multiplying the 
structure (this factor goes to zero in the
$SU(3)_f$ symmetry limit), any uncertainty in the sum rule gets 
enhanced, and hence the results are also not reliable.
Thus we are left with the $\not\!p \not\!q \gamma_5$ structure only.

In order to obtain meson-baryon coupling we need the explicit forms of the 
interpolating currents for the baryons. It is 
well known that there is a continuum number of interpolating currents for the octet baryons. 
In our calculations we will use the following
general forms of baryon currents
\begin{eqnarray}
\eta^{\Sigma^0} &=& \sqrt\frac12 \epsilon^{abc} \left[
\left( u^{aT} C s^b \right) \gamma_5 d^c + t \left( u^{aT} C \gamma_5 s^b
\right) d^c
- \left( s^{aT} C d^b \right) \gamma_5 u^c - t \left( s^{aT} C \gamma_5 d^b
  \right) u^c
  \right]
  \nonumber \\
  \eta^{\Sigma^+} &=& -\frac{1}{\sqrt2} \eta^{\Sigma^0}(d \rightarrow u)
  \nonumber \\ 
  \eta^{\Sigma^-} &=& \frac{1}{\sqrt2} \eta^{\Sigma^0}(u \rightarrow d)
  \nonumber \\
  \eta^p &=& \eta^{\Sigma^+}(s \rightarrow d)
  \nonumber \\
  \eta^n &=&  \eta^{\Sigma^-}(s \rightarrow u)
  \nonumber \\
  \eta^{\Xi^0} &=& \eta^{n}(d \rightarrow s)
  \nonumber \\
  \eta^{\Xi^-} &=&  \eta^{p}(u \rightarrow s)
  \nonumber \\
  \eta^{\Lambda} &=& -\sqrt\frac16 \epsilon^{abc} \left[
  2 \left( u^{aT} C d^b \right) \gamma_5 s^c + 2 t \left( u^{aT} C \gamma_5
  d^b \right) s^c
  +\left( u^{aT} C s^b \right) \gamma_5 d^c 
  \right. \nonumber \\ 
  &+& t\left. \left( u^{aT} C \gamma_5 s^b \right) d^c
  +\left( s^{aT} C d^b \right) \gamma_5 u^c + t \left( s^{aT} C \gamma_5
  d^b \right) u^c
  \right]
  \label{eq3}
  \end{eqnarray}
where $a,~b,~c$ are the color indices, and $t$ is an
arbitrary parameter and $C$ is the charge conjugation operator. 
The Ioffe current  corresponds to the choice
$t=-1$. Note that all currents except the current of $\Lambda$ can be 
obtained from the current of $\Sigma^0$ by simple 
replacements. Recently it has been shown in \cite{R19,R20} that it is 
also possible to obtain the current of 
$\Lambda$ from the current of $\Sigma^0$ through:
\begin{eqnarray}
2 \eta_{\Sigma^0}(d \leftrightarrow s) + \eta_{\Sigma^0}&=& - \sqrt3 \eta_\Lambda 
\nonumber \\
2 \eta_{\Sigma^0}(u \leftrightarrow s) - \eta_{\Sigma^0}&=& - \sqrt3 \eta_\Lambda
\label{sigrel}
\end{eqnarray}
\subsection{Relations Between the Correlation Functions}
Identities presented in Eqs. (\ref{sigrel}) and (\ref{eq3}),  allow us
to write all the correlation functions for the strong coupling constants of the $\pi^{0,\pm}$ and $K^{0,\pm}$
to the baryon octet, in terms of only four functions.
In the $SU(3)_f$ symmetry limit, all these couplings are related using
symmetry arguments. The main power of our approach is that our relations do
not make use of the exact $SU(3)_f$ symmetry and hence can be used to study
various $SU(3)_f$ symmetry violation effects.

Two of these four independent functions can be obtained from a slightly modified form of the correlation function 
$\Pi^{\Sigma^0 \rightarrow \Sigma^0 \pi^0}$.
To start the derivation of the relationships between the various
correlation functions, define:
\begin{eqnarray}
\Pi^{\Sigma^0 \rightarrow \Sigma^0 \pi^0} = g_{\pi \bar u u} \Pi_1(u,d,s)
+ g_{\pi \bar d d} \Pi_1'(u,d,s) + g_{\pi \bar s s} \Pi_2(u,d,s)
\end{eqnarray}
where we formally write the quark content of $\pi^0$ in the form of Eq.
(\ref{eq5}).
For a real pion, we have $g_{\pi \bar u u}  = - g_{\pi \bar d d} =
\frac{1}{\sqrt2}$ and $g_{\pi \bar s s} = 0$.
Hence, essentially $\Pi_1(u,d,s)$, $\Pi'_1(u,d,s)$ and $\Pi_2(u,d,s)$ is the 
contribution to the correlation function
when the pion is emitted from the $u$, $d$, and $s$ quark in $\Sigma^0$ respectively.

Note that  the current for $\Sigma^0$ is symmetric under
the exchange of the $u$ and $d$ quark field operators. Hence the
contribution of emission from the $d$ quark can be obtained from the
contribution of the emission from the $u$ quark by a simple exchange of the
$u$ and $d$ quarks, i.e $\Pi'_1(u,d,s) = \Pi_1(d, u,s)$.
This leaves us with only two independent expressions, i.e. $\Pi_1(u,d,s)$ and
$\Pi_2(u,d,s)$. In the following, we will use the formal notation:
\begin{eqnarray}
\Pi_1(u,d,s) = \langle \bar u u \vert \Sigma^0 \bar{\Sigma^0} \vert 0 \rangle
\nonumber \\
\Pi_2(u,d,s) = \langle \bar s s \vert \Sigma^0 \bar{\Sigma^0} \vert 0 \rangle
\end{eqnarray}

In $\Pi_1$  substituting $d$ instead of $u$, and using the fact that
$\Sigma^0(d \rightarrow u) = \sqrt2 \Sigma^+$, we obtain
\begin{eqnarray}
4 \Pi_1(u,u,s) = 2 \langle \bar u u \vert \Sigma^+ \bar{\Sigma^+}
\vert 0 \rangle
\end{eqnarray}
(The factor $4$ on the left hand side is introduced due to the fact that, since $\Sigma^+$ has two $u$ quark, there are
$4$ ways that the $\pi^0$ can be emitted, but $\Pi_1(u,u,s)$ takes into account only one of these.)
Also noting that since $\Sigma^+$ does not contain any $d$ quark,
\begin{eqnarray}
\Pi^{\Sigma^+ \rightarrow \Sigma^+ \pi^0} &=& g_{\pi \bar u u} 
\langle \bar u u \vert \Sigma^+ \bar{\Sigma^+} \vert 0 \rangle +
g_{\pi \bar s s} \langle \bar s s \vert \Sigma^+ \bar{\Sigma^+} \vert 0
\rangle
\nonumber \\
&=& \sqrt2 \Pi_1(u,u,s)
\end{eqnarray}
Similarly, for $\Sigma^-$, we obtain
\begin{eqnarray}
\Pi^{\Sigma^- \rightarrow \Sigma^-\pi^0} &=& g_{\pi \bar dd} \langle \bar d
d \vert \Sigma^- \bar{\Sigma^-} \vert 0 \rangle + g_{\pi \bar s s} \langle
\bar s s \vert \Sigma^-  \bar{\Sigma^-} \vert 0 \rangle
\nonumber \\
&=&- \sqrt2 \Pi'_1(d,d,s) = - \sqrt2 \Pi_1(d,d,s)
\end{eqnarray}
which concludes derivation of the relations between the couplings of $\pi^0$ to
the $\Sigma$ baryons.

In order to derive relations for the coupling of $\pi^0$ to the proton and
neutron, we need the matrix elements$\langle \bar u u \vert N \bar N \vert
0 \rangle$ and $\langle \bar d d \vert N \bar N \vert 0 \rangle$. In order
to obtain the first matrix element, note that proton current can be
obtained from the $\Sigma^+$ current by replacing the $s$ quark by the $d$
quark. Hence
\begin{eqnarray}
\langle \bar  u u \vert p \bar p \vert 0 \rangle = (\langle \bar u u \vert
\Sigma^+ \bar{\Sigma^+} \vert 0 \rangle)(s \rightarrow d) = \Pi_1(u,u,d)
\end{eqnarray}

In order to obtain $\langle \bar d d \vert p \bar p \vert 0 \rangle$,
$\Pi_2(u,d,s)$ will be needed. First, replacing the $d$ quark, by the $u$
quark, we obtain
\begin{eqnarray}
\Pi_2(u,u,s) = \langle \bar s s \vert \Sigma^+ \bar{\Sigma^+} \vert 0
\rangle
\end{eqnarray}
and next, by replacing the $s$ quark by the $d$ quark, we obtain
\begin{eqnarray}
\Pi_2(u,u,d) = \langle \bar d d  \vert p \bar p \vert 0 \rangle
\end{eqnarray}
Hence we see that,
\begin{eqnarray}
\Pi^{p \rightarrow p \pi^0} &=& g_{\pi \bar u u} 
\langle \bar u u \vert p \bar p \vert 0 \rangle
+ g_{\pi \bar d d} \langle \bar d d  \vert p \bar p \vert 0 \rangle
\nonumber \\
&=&\sqrt2 \Pi_1(u,u,d) - \frac{1}{\sqrt2} \Pi_2(u,u,d)
\end{eqnarray}
Using similar reasoning, one can also derive the following relationships for
the coupling of $\pi^0$ to the nucleon and $\Xi$ baryons:
\begin{eqnarray}
\Pi^{n \rightarrow n \pi^0} &=& \frac{1}{\sqrt2} \Pi_2(d,d,u)
- \sqrt2 \Pi_1(d,d,u)
\nonumber \\
\Pi^{\Xi^0 \rightarrow \Xi^0 \pi^0} &=& \frac{1}{\sqrt2} \Pi_2(s,s,u)
\nonumber \\
\Pi^{\Xi^- \rightarrow \Xi^- \pi^0} &=& -\frac{1}{\sqrt2} \Pi_2(s,s,d)
\end{eqnarray}
which concludes the derivation of the couplings of $\pi^0$ to the baryons
in terms of $\Pi_1(u,d,s)$ and $\Pi_2(u,d,s)$.

Relating the analytical expression of the couplings of the neutral pion to
the coupling of the charged pion is more subtle. To motivate the reasoning,
consider the matrix element $\langle \bar d d \vert \Sigma^0 \bar
{\Sigma^0} \vert 0 \rangle$. This $\Sigma^0$ baryon  contains one of each of 
the $u$, $d$ and $s$ quarks. In this matrix element, it is the $d$ quarks
which emit the final $\bar d d$ and the other $u$ and $d$ quarks act as
spectators. Similarly, in the matrix element
$\langle \bar u d \vert \Sigma^+ \bar {\Sigma^0} \vert 0 \rangle$, the $d$
quark in $\Sigma^0$ and one of the $u$ quarks in $\Sigma^+$ form the state
$\langle \bar u d \vert$ and the other $u$ and $s$ quarks in both the
baryons act as spectator. Thus it is reasonable to expect that $\langle
\bar d d \vert \Sigma^0 \bar {\Sigma^0} \vert 0 \rangle$ and $\langle \bar
u d \vert \Sigma^+ \bar {\Sigma^0} \vert 0 \rangle$ are proportional.
Indeed an explicit calculation of  the correlation
functions showed that
\begin{eqnarray}
\Pi^{\Sigma^0 \rightarrow \Sigma^+ \pi^-} &=& \langle \bar u d \vert \Sigma^+
\bar {\Sigma^0} \vert 0 \rangle 
\nonumber \\
&=& - \sqrt2 \langle \bar d d \vert
\Sigma^0 \bar {\Sigma^0} \vert 0 \rangle = - \sqrt2 \Pi'_1(u,d,s) =
 - \sqrt2 \Pi_1(d,u,s)
\end{eqnarray}
from which, after exchanging $u$ and $d$ quarks, one obtains
\begin{eqnarray}
\Pi^{\Sigma^0 \rightarrow \Sigma^- \pi^+} &=& \langle \bar d u \vert
\Sigma^-\bar {\Sigma^0} \vert 0 \rangle 
\nonumber \\
&=&\sqrt2 \langle \bar u u \vert \Sigma^0 \bar {\Sigma^0} \vert 0 \rangle =
\sqrt2  \Pi_1(u,d,s)
\end{eqnarray}
\newpage
Using a similar reasoning, it is expected that $\langle \bar u u \vert \Xi^0
\bar{\Xi^0} \vert 0 \rangle$ should be proportional to $ \langle \bar d u
\vert \Xi^- \bar{\Xi^0} \vert 0 \rangle$. An explicit calculation showed
that
\begin{eqnarray}
\Pi^{\Xi^0 \rightarrow  \Xi^- \pi^+} = 
\langle \bar d u \vert \Xi^- \bar{\Xi^0} \vert 0 \rangle =
- \sqrt2 \langle \bar u u \vert \Xi^0 \bar{\Xi^0} \vert 0 \rangle =
- \Pi_2(s,s,u)
\end{eqnarray}
and exchanging $u$ and $d$ quarks, we obtain:
\begin{eqnarray}
\Pi^{\Xi^- \rightarrow  \Xi^0 \pi^-} = 
\langle \bar u d \vert \Xi^0 \bar{\Xi^-} \vert 0 \rangle =
- \sqrt2 \langle \bar d d \vert \Xi^0 \bar{\Xi^0} \vert 0 \rangle =
- \Pi_2(s,s,d)
\end{eqnarray}
Using similar arguments, one can show the following relations for the other correlation functions involving
the pion and not involving $\Lambda$:
\begin{eqnarray}
\Pi^{\Sigma^- \rightarrow \Sigma^0 \pi} &=& \sqrt2 \Pi_1(u,d,s)
\nonumber \\
\Pi^{\Sigma^+ \rightarrow \Sigma^0 \pi} &=&- \sqrt2 \Pi'_1(u,d,s) = -\sqrt2 \Pi_1(d,u,s)
\nonumber \\
\Pi^{\Sigma^- \rightarrow n K^-} &=& - \Pi_2(d,d,s)
\nonumber \\
\Pi^{p \rightarrow \Sigma^+ K^0} &=& - \Pi_2(u,u,d)
\nonumber \\
\Pi^{\Sigma^+ \rightarrow p K^0} &=& - \Pi_2(u,u,s)
\nonumber \\
\Pi^{n \rightarrow \Sigma^- K^+} &=& - \Pi_2(d,d,u)
\end{eqnarray}
In order to obtain the expressions for the correlations involving the $\Lambda$ baryon, one uses the relations given in Eq. (\ref{sigrel}). 
Using those relations, one obtains:
\begin{eqnarray}
\Pi^{\Lambda \rightarrow \Lambda \pi^0} &=& \frac{\sqrt2}{3} \left[
\Pi_1(u,s,d) - \Pi_1(d,s,u) + \Pi_2(s,d,u) 
\right. \nonumber \\
&&- \left. \Pi_2(s,u,d) -
\frac12 \Pi_1(u,d,s) + \frac12 \Pi_1(d,u,s) \right]
\nonumber \\
\Pi^{\Lambda \rightarrow \Sigma^0 \pi^0} + 
\Pi^{\Sigma^0 \rightarrow \Lambda \pi^0} &=& \frac{2}{\sqrt6} \left[
\Pi_1(u,s,d) + \Pi_1(d,s,u) 
\right. \nonumber \\
&&- \left. \Pi_2(s,d,u) - \Pi_2(s,u,d) \right]
\nonumber \\
\Pi^{\Xi^- \rightarrow \Sigma^0 K} + \sqrt3 \Pi^{\Xi^- \rightarrow \Lambda
K}&=&  2 \sqrt2 \Pi_1(u,s,d)
\nonumber \\
\Pi^{n \rightarrow \Sigma^0 K} - \sqrt3 \Pi^{n \rightarrow \Lambda K} &=&
2 \sqrt2 \Pi_1(s,d,u)
\nonumber \\
\Pi^{p \rightarrow \Sigma^0 K^+} + \sqrt3 \Pi^{p \rightarrow \Lambda K^+} &=&
- 2 \sqrt2 \Pi_1(s,u,d)
\nonumber \\
- \Pi^{\Xi^0 \rightarrow \Sigma^0 K^0} + \sqrt3 \Pi^{\Xi^0 \rightarrow
  \Lambda K^0} &=& 2 \sqrt2 \Pi_1(d,s,u)
\nonumber \\
\Pi^{\Sigma^0 \rightarrow p K^-} + \sqrt3 \Pi^{\Lambda \rightarrow p K^-}
&=& -2 \sqrt2 \Pi_1(s,u,d)
\nonumber \\
\Pi^{\Sigma^0 \rightarrow n K^0} - \sqrt3 \Pi^{\Lambda \rightarrow n K^0}
&=& 2 \sqrt2 \Pi_1(s,d,u)
\nonumber \\
\Pi^{\Sigma^0 \rightarrow \Xi^0 K^0} - 
\sqrt3 \Pi^{\Lambda \rightarrow \Xi^0 K^0}
&=& - 2 \sqrt2 \Pi_1(d,s,u)
\nonumber \\
\Pi^{\Sigma^0 \rightarrow \Xi^- K^+} + \sqrt3 \Pi^{\Lambda \rightarrow
\Xi^- K^+} &=&  2 \sqrt2 \Pi_1(u,s,d)
\label{eq24}
\end{eqnarray}
As one can see from the relations in Eq. (\ref{eq24}), it is not possible to separate the correlations involving the $\Lambda$ baryon
from expressions involving the $\Sigma^0$ baryon using only the functions $\Pi_1$ and $\Pi_2$. In order to separate these correlation functions,
we will choose two more independent functions:
\begin{eqnarray}
\Pi_3(u,d,s) = -\Pi^{\Sigma^0 \rightarrow \Xi^- K^+} = -\langle u \bar s \vert \Xi^- \bar {\Sigma^0} \vert 0 \rangle 
\end{eqnarray}
and
\begin{eqnarray}
\Pi_4(u,d,s) = -\Pi^{\Xi^- \rightarrow \Sigma^0 K^-} = -\langle s \bar u \vert \Sigma^0 \bar {\Xi^-} \vert 0 \rangle 
\end{eqnarray}
The main motivation for choosing these two correlation functions is that one involves $\Sigma^0$ as an initial state and the other one
involves $\Sigma^0$ as a final state. There are many other possible choices and each one would work equally well. For convention, these two are chosen
in this work.

Using these two correlations, by suitable replacements or exchanges of the quarks, one obtains the following relations:

\begin{eqnarray}
\Pi^{\Xi^0 \rightarrow \Sigma^+ K^-}&=& \Pi^{n
\rightarrow p \pi^-}(s \leftrightarrow d) = -\sqrt2 \Pi_3(s,s,u) 
\nonumber \\
\Pi^{\Xi^- \rightarrow \Sigma^- K^0} &=& \Pi^{\Xi^0 \rightarrow \Sigma^+
K^-}(u \rightarrow d) = -\sqrt2 \Pi_3(s,s,d)
\nonumber \\
\Pi^{\Sigma^+ \rightarrow \Xi^0 K^+} &=& \sqrt2 \Pi^{\Sigma^0
\rightarrow \Xi^- K^+}(d \rightarrow  u) = -\sqrt2 \Pi_3(u,u,s)
\nonumber \\
\Pi^{p \rightarrow n \pi} &=& \Pi^{\Sigma^+ \rightarrow \Xi^0 K^+}(s
\rightarrow d) = -\sqrt2 \Pi_3(u,u,d)
\nonumber \\
\Pi^{n \rightarrow p \pi^-} &=& \Pi^{p \rightarrow n \pi^+}(u
\leftrightarrow d) = -\sqrt2 \Pi_3(d,d,u)
\nonumber \\
\Pi^{\Sigma^0 \rightarrow p K^-} - \sqrt3 \Pi^{\Lambda \rightarrow p K^-}
&=& 2 \Pi^{\Sigma^0 \rightarrow \Xi^- K^+}(s \leftrightarrow u) = - 2 \Pi_3(s,d,u)
\nonumber \\
\Pi^{\Sigma^0 \rightarrow n K^0} + \sqrt3 \Pi^{\Lambda \rightarrow n K^0}
&=& -2 \Pi^{\Sigma^0 \rightarrow \Xi^- K^+}(s \leftrightarrow u,u
\leftrightarrow d) = 2 \Pi_3(s,u,d)
\nonumber \\
\Pi^{\Sigma^0 \rightarrow \Sigma^- \pi^+} + \sqrt3 \Pi^{\Lambda \rightarrow
\Sigma^- \pi^+} &=& 2 \Pi^{\Sigma^0 \rightarrow \Xi^- K^+} (s
\leftrightarrow d) = - 2 \Pi_3(u,s,d)
\nonumber \\
\Pi^{\Lambda \rightarrow \Sigma^+ \pi^-} &=&  \Pi^{\Lambda \rightarrow
\Sigma^- \pi^+}(u \leftrightarrow d) 
\nonumber \\
\Pi^{\Sigma^- \rightarrow \Xi^- K} &=& \Pi^{\Sigma^+ \rightarrow \Xi^0 K}(u
\rightarrow d) = -\sqrt2 \Pi_3(d,d,s);
\nonumber \\
\Pi^{\Sigma^0 \rightarrow \Xi^0 K^0} &=& -\Pi^{\Sigma^0 \rightarrow \Xi^- K^+}
(d \leftrightarrow u) = \Pi_3(d,u,s)
\nonumber \\
\Pi^{\Xi^0 \rightarrow \Sigma^0 K^0} &=& - \Pi^{\Xi^- \rightarrow \Sigma^0
K^+}(u \leftrightarrow d) = \Pi_4(d,u,s)
\nonumber \\
\Pi^{p \rightarrow \Sigma^0 K^+} - \sqrt3 \Pi^{p \rightarrow \Lambda K^+}
&=& 2 \Pi^{\Xi^- \rightarrow \Sigma^0 K^-}(u \leftrightarrow s) = -2 \Pi_4(s,d,u)
\nonumber \\
\Pi^{n \rightarrow \Sigma^0 K^0} + \sqrt3 \Pi^{n \rightarrow \Lambda K^0}
&=& -2 \Pi^{\Xi^- \rightarrow \Sigma^0 K^-}(u \leftrightarrow s, u
\leftrightarrow d) = 2 \Pi_4(s,u,d)
\nonumber \\
\Pi^{\Sigma^- \rightarrow \Sigma^0 \pi^-} + \sqrt3 \Pi^{\Sigma^- \rightarrow
\Lambda \pi^-} &=&  2 \Pi^{\Xi^- \rightarrow \Sigma^0 K^-}(d \leftrightarrow
s) = -2 \Pi_4(u,s,d)
\nonumber \\
\Pi^{\Sigma^+ \rightarrow \Lambda \pi^+} &=& \Pi^{\Sigma^- \rightarrow \Lambda \pi^-}(u \leftrightarrow d)
\end{eqnarray}
Hence we conclude our claim that all 45 strong coupling constants of the pions and the kaon to the octet baryons can be expressed
in terms of only 4 independent function without using any flavor symmetry. In Appendix A, we present each one of the correlation functions in terms
of the four functions $\Pi_i(u,d,s)$, ($i=1,~2,~3,~4$). In this work, we will work in the exact isosymmetry limit. In this limit, not all of the
coupling constants are independent. In Appendix B, we present the relationships between the coupling constants in this limit.
\subsection{Expressions for the Functions $\Pi_i$}

In the previous subsection, we have shown that all the correlation functions can be expressed in terms of only four analytical functions which
can be obtained from the correlation functions for the transitions $\Sigma^0 \rightarrow \Sigma^0 \pi^0$, $\Sigma^0 \rightarrow \Xi^0 K^0$ and 
$\Xi^0 \rightarrow \Sigma^0  K^0$. Hence it is enough to evaluate only the correlation functions for these transitions only.

In the large Euclidean momentum $-p_1^2 \rightarrow \infty$ and $-p_2^2 \rightarrow \infty$ region, the
correlators can be calculated using the OPE. For this calculation, the propagators of the light quarks
and matrix elements of the form $\langle {\cal M} \vert \bar q(x_1) \Gamma q'(x_2) \vert 0 \rangle$ where ${\cal M}=K^0$ or $\pi^0$, and
$\Gamma$ is a
member of the Dirac basis of the gamma matrices are needed. In order to study the $SU(3)_f$ violation effects, we have expressed
the light quark propagator up to linear order in $m_q$ and then for numerical analysis set $m_u=m_d=0$ and $m_s \neq 0$.
The matrix elements $\langle {\cal M} \vert \bar q(x_1) \Gamma q'(x_2) \vert 0 \rangle$ can be written in terms of the meson
light cone
distribution amplitudes. The explicit forms of these matrix elements are given in \cite{R21,R22}:
\begin{eqnarray}
\langle {\cal M}(p)| \bar q(x) \gamma_\mu \gamma_5 q(0)| 0 \rangle &=& -i f_{\cal M} p_\mu  \int_0^1 du  e^{i \bar u p x} 
	\left( \varphi_{\cal M}(u) + \frac{1}{16} m_{\cal M}^2 x^2 {\mathbb A}(u) \right)
\nonumber \\
	&-& \frac{i}{2} f_{\cal M} m_{\cal M}^2 \frac{x_\mu}{px} \int_0^1 du e^{i \bar u px} {\mathbb B}(u) 
\nonumber \\
\langle {\cal M}(p)| \bar q(x) i \gamma_5 q(0)| 0 \rangle &=& \mu_{\cal M} \int_0^1 du e^{i \bar u px} \varphi_P(u)
\nonumber \\
\langle {\cal M}(p)| \bar q(x) \sigma_{\alpha \beta} \gamma_5 q(0)| 0 \rangle &=& 
\frac{i}{6} \mu_{\cal M} \left( 1 - \tilde \mu_{\cal M}^2 \right) \left( p_\alpha x_\beta - p_\beta x_\alpha\right)
	\int_0^1 du e^{i \bar u px} \varphi_\sigma(u)
\nonumber \\
\langle {\cal M}(p)| \bar q(x) \sigma_{\mu \nu} \gamma_5 g_s G_{\alpha \beta}(v x) q(0)| 0 \rangle &=&
	i \mu_{\cal M} \left[
		p_\alpha p_\mu \left( g_{\nu \beta} - \frac{1}{px}(p_\nu x_\beta + p_\beta x_\nu) \right) 
\right. \nonumber \\
	&-&	p_\alpha p_\nu \left( g_{\mu \beta} - \frac{1}{px}(p_\mu x_\beta + p_\beta x_\mu) \right) 
\nonumber \\
	&-&	p_\beta p_\mu \left( g_{\nu \alpha} - \frac{1}{px}(p_\nu x_\alpha + p_\alpha x_\nu) \right)
\nonumber \\ 
	&+&	p_\beta p_\nu \left. \left( g_{\mu \alpha} - \frac{1}{px}(p_\mu x_\alpha + p_\alpha x_\mu) \right)
		\right]
\nonumber \\
	&\times& \int {\cal D} \alpha e^{i (\alpha_{\bar q} + v \alpha_g) px} {\cal T}(\alpha_i)
\nonumber \\
\langle {\cal M}(p)| \bar q(x) \gamma_\mu \gamma_5 g_s G_{\alpha \beta} (v x) q(0)| 0 \rangle &=& 
	p_\mu (p_\alpha x_\beta - p_\beta x_\alpha) \frac{1}{px} f_{\cal M} m_{\cal M}^2 
		\int {\cal D}\alpha e^{i (\alpha_{\bar q} + v \alpha_g) px} {\cal A}_\parallel (\alpha_i)
\nonumber \\
	&+& \left[
		p_\beta \left( g_{\mu \alpha} - \frac{1}{px}(p_\mu x_\alpha + p_\alpha x_\mu) \right) \right.
\nonumber \\
	&-& 	p_\alpha \left. \left(g_{\mu \beta}  - \frac{1}{px}(p_\mu x_\beta + p_\beta x_\mu) \right) \right]
	f_{\cal M} m_{\cal M}^2
\nonumber \\
	&\times& \int {\cal D}\alpha e^{i (\alpha_{\bar q} + v \alpha _g) p x} {\cal A}_\perp(\alpha_i)
\nonumber \\
\langle {\cal M}(p)| \bar q(x) \gamma_\mu i g_s G_{\alpha \beta} (v x) q(0)| 0 \rangle &=& 
	p_\mu (p_\alpha x_\beta - p_\beta x_\alpha) \frac{1}{px} f_{\cal M} m_{\cal M}^2 
		\int {\cal D}\alpha e^{i (\alpha_{\bar q} + v \alpha_g) px} {\cal V}_\parallel (\alpha_i)
\nonumber \\
	&+& \left[
		p_\beta \left( g_{\mu \alpha} - \frac{1}{px}(p_\mu x_\alpha + p_\alpha x_\mu) \right) \right.
\nonumber \\
	&-& 	p_\alpha \left. \left(g_{\mu \beta}  - \frac{1}{px}(p_\mu x_\beta + p_\beta x_\mu) \right) \right]
	f_{\cal M} m_{\cal M}^2
\nonumber \\
	&\times& \int {\cal D}\alpha e^{i (\alpha_{\bar q} + v \alpha _g) p x} {\cal V}_\perp(\alpha_i)
\end{eqnarray}
where $\mu_{\cal M} = f_{\cal M} \frac{m_{\cal M}^2}{m_{q_1} + m_{q_2}},$ 
$\tilde \mu_{\cal M} = \frac{m_{\cal M}}{m_{q_1} + m_{q_2}}$, where
$q_1$ and $q_2$ are the quarks in the meson ${\cal M}$,
${\cal D}\alpha = d\alpha_{\bar q} d\alpha_q d\alpha_g \delta(1-\alpha_{\bar q} - \alpha_q - \alpha_g)$, and
the functions $\varphi_{\cal M}(u),$ $\mathbb A(u),$ $\mathbb B(u),$ $\varphi_P(u),$ $\varphi_\sigma(u),$ 
${\cal T}(\alpha_i),$ ${\cal A}_\perp(\alpha_i),$ ${\cal A}_\parallel(\alpha_i),$ ${\cal V}_\perp(\alpha_i)$ and ${\cal V}_\parallel(\alpha_i)$
are functions of definite twist and their expressions will be given in the Numerical Analysis section.

For the explicit form
of the light quark propagator, we have used the expression:
\begin{eqnarray}
S_q(x) &=& \frac{i \not\!x}{2\pi^2 x^4} - \frac{m_q}{4 \pi^2 x^2} - \frac{\langle \bar q q \rangle}{12}
\left(1 - i \frac{m_q}{4} \not\!x \right) - \frac{x^2}{192} m_0^2 \langle \bar q q \rangle 
\left( 1 - i \frac{m_q}{6}\not\!x \right) 
\nonumber \\ &&
 - i g_s \int_0^1 du \left[\frac{\not\!x}{16 \pi^2 x^2} G_{\mu \nu} (ux) \sigma_{\mu \nu} - u x^\mu
 G_{\mu \nu} (ux) \gamma^\nu \frac{i}{4 \pi^2 x^2} 
 \right. \nonumber \\ && \left.
 - i \frac{m_q}{32 \pi^2} G_{\mu \nu} \sigma^{\mu
 \nu} \left( \ln \left( \frac{-x^2 \Lambda^2}{4} \right) +
 2 \gamma_E \right) \right]
\label{prop}
\end{eqnarray}
where $\gamma_E$ is the Euler Constant, $\gamma_E \simeq 0.577$ and $\Lambda$ is a scale that separates the 
long and short distance physics.
For numerical analysis, we used the value $\Lambda=300~MeV$.

Using the explicit expressions of the full propagator of the light quark, Eq. (\ref{prop}), and the meson wave functions, and 
separating the coefficient of the structure $\not\!p \not\! q \gamma_5$ one can obtain the theoretical expression for the 
correlation function in terms of a few condensates, distribution amplitudes of the mesons and the QCD parameter.

Having the explicit expression for the correlator from QCD, 
QCD sum rules is obtained by 
applying Borel transformations on the variables $p_1^2$ and $p_2^2 = (p_1+q)^2$ in order to suppress the contribution of
the higher states and the continuum (for details see e.g. \cite{R23,R24,R25}). Then equating the final results 
from phenomenological and corresponding parts we arrive at the sum rules for the corresponding kaon baryon couplings. Our final results for 
the four analytical functions $\Pi_i$, $i=1,~2,~3,$ and $4$ (for the structure $\not\!p \not\!q \gamma_5$) are:
\begin{eqnarray}
&&\Pi_1(u,d,s) =
\nonumber \\
&& \frac{f_{\cal M}}{64 \pi^2} M^4 \left( m_s (1-t)^2 - m_d (1-t^2) \right) 
i_2(\phi_{\cal M})
\nonumber \\
&-& \frac{\mu_{\cal M}}{64 \pi^2} M^4 \left(1 - \tilde \mu_{\cal M}^2 \right)
(1-t^2) i_2(\phi_\sigma)
\nonumber \\
&-& \frac{f_{\cal M}}{32 \pi^2} m_{\cal M}^2 M^2 \left( m_s (1-t)^2 + 3 m_d (1-t^2) 
\right) i_1({\cal V},1) \left( \gamma_E - \ln \frac{M^2}{\Lambda^2} \right)
\nonumber \\
&+& \frac{f_{\cal M}}{16 \pi^2} m_{\cal M}^2 M^2 \left( m_s (1-t)^2 - m_d (1-t^2) \right)
i_1({\cal V}_\perp,1) \left( \gamma_E - \ln \frac{M^2}{\Lambda^2} \right)
\nonumber \\
&+& \frac{f_{\cal M}}{768 \pi^2} \langle g_s^2 GG \rangle \left( m_s (1-t)^2 - m_d (1-t^2) 
\right) i_2(\phi_{\cal M}) \left( \gamma_E - \ln \frac{M^2}{\Lambda^2} \right)
\nonumber \\
&-& \frac{m_0^2+ 2 M^2}{3456 M^6} \mu_{\cal M} \left( 1- \tilde \mu_{\cal M}^2 \right)
\langle g_s^2 GG \rangle \left( m_d \sp + m_s \dd \right) 
(3 + 2 t + 3 t^2 ) i_2(\phi_\sigma)
\nonumber \\
&-& \frac{m_0^2}{648 M^2} \left(1 - \tilde \mu_{\cal M}^2 \right) \mu_{\cal M} 
(m_s \dd + m_d \sp) (5 + 4 t + 5 t^2) i_2(\phi_\sigma)
\nonumber \\
&-& \frac{f_{\cal M}}{3072 \pi^2 M^2} \GG m_{\cal M}^2 
\left( m_s (1-t)^2 - m_d (1-t^2) \right) 
\left( 4 i_1({\cal A}_\parallel, 1-2v) - i_2({\mathbb A}) \right)
\nonumber \\
&+& \frac{f_{\cal M}}{768\pi^2 M^2} \GG m_{\cal M}^2 
\left( m_s (1-t)^2 + m_d (1-t^2) \right)  i_1({\cal V}_\parallel,1)
\nonumber \\
&-& \frac{f_{\cal M}}{1152 \pi^2} \GG \left( m_s (1-t)^2 - m_d (1-t^2) \right)
  i_2(\varphi_{\cal M})
\nonumber \\
&-& \frac{f_{\cal M}}{128 \pi^2} m_{\cal M}^2 M^2 \left( m_s (1-t)^2 - m_d (1-t^2)
\right) i_2(\mathbb A)
\nonumber \\
&-& \frac{f_{\cal M}}{24} M^2 \left( \sp (1-t)^2 - \dd (1-t^2) \right)
i_1(\varphi_{\cal M})
\nonumber \\
&+& \frac{f_{\cal M}}{32 \pi^2} m_{\cal M}^2 M^2 \left( m_s (1-t)^2 - m_d (1-t^2)
\right) i_1({\cal A}_\parallel,1-2 v)
\nonumber \\
&-& \frac{f_{\cal M}}{16 \pi^2} m_{\cal M}^2 M^2 \left( m_s (1-t)^2 + 2 m_d (1-t^2)
\right) i_1 ({\cal V}_\parallel,1)
\nonumber \\
&+& \frac{f_{\cal M}}{16 \pi^2} m_{\cal M}^2 M^2 \left( m_s (1-t)^2 - m_d (1-t^2)
\right) i_1({\cal V}_\perp,1)
\nonumber \\
&+& \frac{\mu_{\cal M}}{144} \left( 1- \tilde \mu_{\cal M}^2 \right)
\left[ \dd \left( - 3 m_d (1-t^2) - 2 m_s (3 + 2 t + 3 t^2) \right) 
\right. \nonumber \\ 
&-& \left.
\sp \left( 3 m_s (1-t^2) + m_d ( 6 + 4 t + 6 t^2 ) \right) \right]
i_2(\phi_\sigma)
\nonumber \\
&+& \frac{f_{\cal M}}{432} m_0^2 \left( 3 \sp (1-t)^2 - 2 \dd (1-t)^2 \right)
i_2(\varphi_{\cal M})
\nonumber \\
&+& \frac{f_{\cal M}}{96} m_{\cal M}^2 \left( \sp (1-t)^2 - \dd (1-t^2) \right)
i_2(\mathbb A)
\nonumber \\
&-& \frac{\mu_{\cal M}}{48} (1-t^2) (\dd m_d - \sp m_s) i_1'({\cal
T},1-2v)
\nonumber \\
&-& \frac{f_{\cal M}}{24} m_{\cal M}^2 \left( \sp (1-t)^2 - \dd (1-t^2) \right)
i_1({\cal A}_\parallel,1-2v)
\nonumber \\
&+& \frac{f_{\cal M}}{24} m_{\cal M}^2 \left( \dd (1-t^2) + \sp (1-t)^2 \right)
i_1({\cal V}_\parallel,1)
\label{pi1}
\end{eqnarray}
\begin{eqnarray}
&&\Pi_2(u,d,s) = 
\nonumber \\
&-& \frac{M^4}{192 \pi^2} \mu_{\cal M} \left( 1 - \tilde \mu_{\cal M} \right) (1-t)^2 i_2(\phi_\sigma)
\nonumber \\
&+& \frac{f_{\cal M}}{64 \pi^2} M^4 (m_u + m_d) (1-t^2) i_2(\varphi_{\cal M})
\nonumber \\
&-& \frac{M^4}{32 \pi^2} \mu_{\cal M} (1-t)^2 i_1'({\cal T},1-2v)
\nonumber \\
&+& \frac{f_{\cal M}}{768 \pi^2} (m_u + m_d) (1-t^2) \GG \left( \gamma_E - \ln \frac{M^2}{\Lambda^2} \right) i_2(\varphi_{\cal M})
\nonumber \\
&-& \frac{f_{\cal M}}{32 \pi^2} m_{\cal M}^2 (m_u + m_d) M^2 (1-t^2) \left( \gamma_E - \ln \frac{M^2}{\Lambda^2} \right)
i_1(3 {\cal V}_\parallel - 2 {\cal V}_\perp,1)
\nonumber \\
&+&\frac{\mu_{\cal M}}{3456 M^6} (1 - \tilde \mu^2) (m_0^2+ 2 M^2) \GG (1+t)^2 (\dd m_u + m_d \uu) i_2(\phi_\sigma)
\nonumber \\
&-& \frac{m_0^2}{1296 M^2} \mu_{\cal M} (1- \tilde \mu_{\cal M}^2) (1+t)^2 (\dd m_u + m_d \uu) i_2(\phi_\sigma)
\nonumber \\
&-& \frac{f_{\cal M}}{3072 \pi^2 M^2} m_{\cal M}^2 \GG (m_u +  m_d) (1-t^2) \left(
4 i_1({\cal A}_\parallel, 1- 2 v) + 4 i_1({\cal V}_\parallel,1) -
i_2(\mathbb A) \right)
\nonumber \\
&-& \frac{f_{\cal M}}{24} M^2 (1-t^2) (\dd + \uu) i_2(\phi_{\cal M})
\nonumber \\
&+& \frac{f_{\cal M}}{128 \pi^2} m_{\cal M}^2 (m_u + m_d) M^2 (1-t^2)
\left( 4 i_1({\cal A}_\parallel,1-2 v) - 8 i_1({\cal V}_\parallel - {\cal
V}_\perp,1) - i_2(\mathbb A) \right)
\nonumber \\
&+& \frac{\mu_{\cal M}}{144} (1- \tilde \mu_{\cal M}^2) \left[
\dd \left( m_d (1-t^2) + 2 m_u (1+t)^2 \right) \right.
\nonumber \\ 
&+& \left.
\uu \left( m_u (1-t^2) - 2 m_d (1+t)^2 \right) \right] i_2(\varphi_\sigma)
\nonumber \\
&+& \frac{f_{\cal M}}{864} (1-t^2) (\dd + \uu) \left( 9 m_{\cal M}^2 i_2(\mathbb A) +
10 m_0^2 i_2(\varphi_{\cal M}) \right)
\nonumber \\
&-& \frac{f_{\cal M}}{1152 \pi^2} \GG (m_u + m_d) (1-t^2) i_2(\varphi_{\cal M})
\nonumber \\
&-& \frac{f_{\cal M}}{24} m_{\cal M}^2 (1-t^2) (\dd + \uu) \left(
i_1({\cal A}_\parallel,1-2v) + i_1({\cal V}_\parallel,1) \right)
\nonumber \\ 
&-& \frac{\mu_{\cal M}}{48} (1-t)^2 (\dd m_d + \uu m_u) i_1'({\cal T},1-2v)
\label{pi2}
\end{eqnarray}
\begin{eqnarray}
&&\sqrt2 \Pi_3(u,d,s) = 
\nonumber \\
&& \frac{f_{\cal M}}{64 \pi^2} M^4 (m_d+m_s) 
\left(1 + 2 t - 3 t^2 \right)
\nonumber \\ 
&+&\frac{1}{192 \pi^2} \left(1 - \tilde \mu_{\cal M}^2 \right) M^4 
\left(5 + 2 t - 7 t^2 \right)
\nonumber \\
&-& \frac{1}{32\pi^2} \mu_{\cal M} (1-t)^2 i_1'({\cal T},1-2v)
\nonumber \\
&+& \frac{f_{\cal M}}{768 \pi^2} \GG (m_d + m_s) \left( 1 + 2 t - 3 t^2 \right)
i_2(\varphi_{\cal M}) \left( \gamma_E - \ln \frac{M^2}{\Lambda^2} - \frac23\right)
\nonumber \\
&-& \frac{f_{\cal M}}{32 \pi^2} m_{\cal M}^2 M^2 (1-t)^2 \left( (m_d - m_s)
i_1({\cal A}_\parallel,1) - (m_d + m_s) i_1({\cal V}_\parallel,1) \right)
\left( \gamma_E - \ln \frac{M^2}{\Lambda^2} \right)
\nonumber \\
&-& \frac{f_{\cal M}}{16 \pi^2} m_{\cal M}^2 M^2 (1 + 2 t - 3 t^2)
\left( (m_d-m_s) i_1({\cal A}_\perp,1) - (m_d+m_s) i_1({\cal V}_\perp,1)
\right)
\left( \gamma_E - \ln \frac{M^2}{\Lambda^2} \right)
\nonumber \\
&+& \frac{\mu_{\cal M}}{3456 M^6} (1 - \tilde \mu_{\cal M}^2) 
\left( m_0^2+2 M^2 \right) \GG
\left( 7 + 6 t + 7 t^2 \right) \left(\dd m_s + \sp m_d \right) 
i_2(\varphi_\sigma)
\nonumber \\
&+& \frac{f_{\cal M}}{3072 M^2 \pi^2} \GG m_{\cal M}^2 (m_d+m_s) (1 + 2 t - 3 t^2)
i_2({\mathbb A})
\nonumber \\
&-& \frac{f_{\cal M}}{768 M^2 \pi^2} \GG m_{\cal M}^2 (m_d + m_s) (1-t) 
\left( (1+3t) i_1({\cal A}_\parallel,1-2v) + (3 + t) i_1({\cal
V}_\parallel,1) \right)
\nonumber \\
&+& \frac{f_{\cal M}}{384 M^2 \pi^2} m_{\cal M}^2 \GG (m_d-m_s)(1-t)
\left( (1+3t) i_1({\cal A}_\perp,1) + (3+t) i_1({\cal V}_\perp,1-2v)
\right)
\nonumber \\
&+& \frac{m_0^2}{1296 M^2} \left( 1 - \tilde \mu_{\cal M}^2 \right) \mu_{\cal M} 
\left(\dd m_s + \sp m_d \right) \left( 19 + 14 t + 19 t^2 \right)
i_2(\phi_\sigma)
\nonumber \\
&-& \frac{f_{\cal M}}{384 \pi^2} \left(1+2t-3t^2 \right)
\left(3 m_{\cal M}^2 (m_d+m_s) i_2({\mathbb A}) + 16 \pi^2 (\dd + \sp)
i_2(\varphi_{\cal M}) \right)
\nonumber \\
&-& \frac{f_{\cal M}}{16 \pi^2} m_{\cal M}^2 (m_d+m_s) M^2 (1-t) \left(
(1+3t) i_1({\cal A}_\parallel,v) - 2 i_1({\cal V}_\parallel,1) \right)
\nonumber \\
&+& \frac{f_{\cal M}}{16 \pi^2} m_{\cal M}^2 M^2 (1-t) \left( 
(m_s(1+t) + 2 m_d t) i_1({\cal A}_\parallel,1) + 2 (2 m_s(1+t) - m_d(1-t))
i_1({\cal V}_\perp,1) \right)
\nonumber \\
&-& \frac{f_{\cal M}}{8 \pi^2} m_{\cal M}^2 (m_d-m_s) M^2 (1-t) \left(
(1+3t) i_1({\cal A}_\perp,1) - (3 +t) i_1({\cal V}_\perp,v) \right)
\nonumber \\
&-& \frac{\mu_{\cal M}}{144} (1 - \tilde \mu_{\cal M}^2) (\dd m_d + \sp m_s)
(-5 - 2 t + 7 t^2) i_2(\varphi_\sigma)
\nonumber \\
&+& \frac{\mu_{\cal M}}{72} (1 - \tilde \mu_{\cal M}^2) (\dd m_s + \sp m_d)
(7 + 6 t + 7 t^2) i_2(\varphi_\sigma)
\nonumber \\
&+& \frac{f_{\cal M}}{864} (\dd + \sp) \left(
9 m_{\cal M}^2 (1 + 2 t - 3 t^2) i_2({\mathbb A}) + 4 m_0^2 (2 + 3 t - 5 t^2)
i_2(\varphi_{\cal M}) \right)
\nonumber \\
&-& \frac{\mu_{\cal M}}{24} (1-t) \left( - 2 m_s \sp t + \dd m_d (1+t) \right)
i_1'({\cal T},1)
\nonumber \\
&+& \frac{\mu_{\cal M}}{24} (\dd m_d + \sp m_s) (1-t)^2 i_1'({\cal T},v)
\nonumber \\
&-& \frac{f_{\cal M}}{24} m_{\cal M}^2 (\dd + \sp) (1-t) \left[
(1+3t) i_1({\cal A}_\parallel,1-2v) + (3 +t) i_1({\cal V}_\parallel,v)
\right]
\nonumber \\
&+& \frac{f_{\cal M}}{12} m_{\cal M}^2 (\dd - \sp) (1-t) \left[
(1+3t) i_1({\cal A}_\perp) + (3 +t) i_1({\cal V}_\perp,1-2v) \right]
\label{pi3}
\end{eqnarray}
\begin{eqnarray}
&&\sqrt2 \Pi_4(u,d,s) = \sqrt2 \Pi_3(u,d,s) 
\nonumber \\
&-& \frac{f_{\cal M}}{16 \pi^2} m_{\cal M}^2 (m_d- m_s) M^2 (1-t)
\left[ (-1+t) i_1({\cal A}_\parallel,1) - 2 (1 + 3 t) i_1({\cal A}_\perp,1)
\right] \times
\nonumber \\
&& \left( \gamma_E - \ln \frac{M^2}{\Lambda^2} \right)
\nonumber \\
&-& \frac{f_{\cal M}}{192 M^2 \pi^2} m_{\cal M}^2 \GG (m_d-m_s) (1-t) 
\left[(1+3t) i_1({\cal A}_\perp,1) + (3 + t) i_1({\cal V}_\perp,1-2v)
\right]
\nonumber \\
&-& \frac{f_{\cal M}}{16 \pi^2} m_{\cal M}^2 (m_d-m_s) M^2 (1-t)
\left[ (-1+t) i_1({\cal A}_\parallel,1) - 4 (1+3t) i_1({\cal A}_\perp,1)
\right]
\nonumber \\
&-& \frac{f_{\cal M}}{8 \pi^2} m_{\cal M}^2 (m_d-m_s) M^2 (-3 + 2 t + t^2)
i_1({\cal V}_\perp,1-2v) 
\nonumber \\
&-& \frac{\mu_{\cal M}}{24} \left( \dd m_d - \sp m_s \right) (-1 - 2 t + 3 t^2)
i_1'({\cal T},1)
\nonumber \\
&-& \frac{f_{\cal M}}{6} m_{\cal M}^2 (\dd - \sp) (1-t) \left[
(1+3t) i_1({\cal A}_\perp,1) + (3 +t) i_1({\cal V}_\perp,1-2v) \right]
\label{pi4}
\end{eqnarray}
where $\cal M=\pi$ or $K$.

The functions $i_n, ~i_n',~ \tilde i_n, ~\tilde {\tilde i}_n$  are defined as:
\begin{eqnarray}
i_1(\varphi,f(v)) &=& \int {\cal D}\alpha_i \int_0^1 dv \varphi(\alpha_{\bar q},\alpha_q,\alpha_g) f(v)
\delta(k - \bar u_0)
\nonumber \\
i_1'(\varphi,f(v)) &=& \int {\cal D}\alpha_i \int_0^1 dv \varphi(\alpha_{\bar q},\alpha_q,\alpha_g)f(v)
\delta'(k - \bar u_0)
\nonumber \\
\tilde i_1(\varphi,f(v)) &=& \int{\cal D}\alpha_i \int_0^1 dv \int_0^{\alpha_{\bar q} + v \alpha_g} dk
\varphi(\alpha_{\bar q},\alpha_q,\alpha_g) f(v) \delta(k-\bar u_0)
\nonumber \\
i_2(f) &=& f(u_0)
\end{eqnarray}
where $k= \alpha_{\bar q} + v \alpha_g$ when there is no integration over $k$, $u_0 = \frac{M_1^2}{M_1^2 + M_2^2}$.

For obtaining the meson baryon couplings, we also need to know the overlap amplitude of the hadrons. The overlap amplitudes 
can be obtained from the mass sum rules and their expressions can be found in \cite{R23,R24}. As we noted that the other
currents can be obtained from $\Sigma^0$ current and therefore, we present result for the residues of this current
only:
\begin{eqnarray}
\lambda_{\Sigma^0}^2 e^{-\frac{m_\Sigma^2}{M^2}} &=&
\frac{M^6}{1024 \pi^2} (5+2t+5t^2) - \frac{m_0^2}{96 M^2}(-1+t)^2 \uu \dd 
\nonumber \\ &&
- \frac{m_0^2}{16 M^2} (-1+t^2) \sp (\dd + \uu)
\nonumber \\ &&
+\frac{3}{128 m_0^2}m_0^2 (-1+t^2) \left[ m_s (\dd + \uu)  + \sp (m_u + m_d) \right]
\nonumber \\ &&
- \frac{1}{64 \pi^2}(-1+ t)^2 \left(\dd m_u + \uu m_d \right) M^2 
\nonumber \\ &&
- \frac{3}{64 \pi^2} (-1+t^2) \left( m_s (\dd + \uu)  + \sp (m_u + m_d ) \right)  M^2 
\nonumber \\ &&
+\frac{1}{128 \pi^2} (5 + 2 t + 5 t^2) \left( \uu m_u + \dd m_d + \sp m_s \right)
\nonumber \\ &&
+ \frac{1}{24} \left[ 3 \sp (\dd + \uu) \left(-1+t^2\right) + (-1+t)^2 \uu \dd  \right]
\nonumber \\ &&
+ \frac{m_0^2}{256 \pi^2} (-1+t)^2 \left(m_d \uu + m_u \dd \right)
\nonumber \\ &&
+ \frac{m_0^2}{256 \pi^2} (-1+t^2) \left[ 13 m_s (\uu + \dd )+ 11 \sp (m_d + m_u) \right]
\nonumber \\ &&
- \frac{m_0^2}{192 \pi^2} (1+t+t^2) \left(\uu m_u +  \dd m_d -2 m_s \sp \right)
\label{masssigma0}
\end{eqnarray}
Note that from the mass sum rules, one can only extract the square of the residue, and not the sign. We have chosen our
sign convention in defining the currents such that in the $SU(3)_f$ symmetry limit, the signs correctly reproduce
the $F$ and $D$ expressions (see for example \cite{R26}).

The contribution of the continuum to the sum rules obtained from Eqs. (\ref{pi1}-\ref{pi4}) and to the mass sum rules Eq. (\ref{masssigma0})
are subtracted using the replacements:
\begin{eqnarray}
M^{2n} \rightarrow \frac{1}{\Gamma(n)} \int_{q^2}^{s_0} ds s^{n-1} e^{-\frac{s}{M^2}},~~~ n > 0
\nonumber \\
M^{2n} \ln \frac{M^2}{\Lambda^2} \rightarrow \frac{1}{\Gamma(n)} \int_{q^2}^{s_0} ds s^{n-1} \left(\ln \frac{s}{\Lambda^2} - \psi(n) \right)
\label{subt}
\end{eqnarray}
where $q^2=0$ in the mass sum rules and $q^2=m_{\cal M}^2$ in the sum rules for
the coupling constants
since the meson is real, $\psi(n)$ is the digamma functions:
\begin{eqnarray}
\psi(x) = \frac{d}{dx} \ln \Gamma[x]
\end{eqnarray}
The subtraction scheme, Eq. (\ref{subt}), corresponds to taking a
triangular domain in the double dispersion relation for the coupling
constant outside of which we use the quark hadron duality to subtract the
contributions of the higher states and continuum.

\section{Numerical Analysis}
In this section we present our numerical results for the sum rules obtained in the previous section for the meson baryon 
coupling constants. The meson baryon coupling constants are physically measurable quantities, they should be independent
on the auxiliary Borel parameter $M^2$, the continuum threshold $s_0$, and the parameter $t$. Therefore we need to find regions
of these parameters where meson baryon couplings are independent of them.

From the sum rules, one sees that the main input parameters are the meson wave functions. In our calculations, we will use 
the following forms of the meson wave functions \cite{R21,R22}
\begin{eqnarray}
\phi_{\cal M}(u) &=& 6 u \bar u \left( 1 + a_1^{\cal M} C_1(2 u -1) + a_2^{\cal M} C_2^{3 \over 2}(2 u - 1) \right) 
\nonumber \\
{\cal T}(\alpha_i) &=& 360 \eta_3 \alpha_{\bar q} \alpha_q \alpha_g^2 \left( 1 + w_3 \frac12 (7 \alpha_g-3) \right)
\nonumber \\
\phi_P(u) &=& 1 + \left( 30 \eta_3 - \frac{5}{2} \frac{1}{\mu_{\cal M}^2}\right) C_2^{1 \over 2}(2 u - 1) 
\nonumber \\ 
&+&	\left( -3 \eta_3 w_3  - \frac{27}{20} \frac{1}{\mu_{\cal M}^2} - \frac{81}{10} \frac{1}{\mu_{\cal M}^2} a_2^{\cal M} \right) C_4^{1\over2}(2u-1)
\nonumber \\
\phi_\sigma(u) &=& 6 u \bar u \left[ 1 + \left(5 \eta_3 - \frac12 \eta_3 w_3 - \frac{7}{20}  \mu_{\cal M}^2 - \frac{3}{5} \mu_{\cal M}^2 a_2^{\cal M} \right)
C_2^{3\over2}(2u-1) \right]
\nonumber \\
{\cal V}_\parallel(\alpha_i) &=& 120 \alpha_q \alpha_{\bar q} \alpha_g \left( v_{00} + v_{10} (3 \alpha_g -1) \right)
\nonumber \\
{\cal A}_\parallel(\alpha_i) &=& 120 \alpha_q \alpha_{\bar q} \alpha_g \left( 0 + a_{10} (\alpha_q - \alpha_{\bar q}) \right)
\nonumber \\
{\cal V}_\perp (\alpha_i) &=& - 30 \alpha_g^2\left[ h_{00}(1-\alpha_g) + h_{01} (\alpha_g(1-\alpha_g)- 6 \alpha_q \alpha_{\bar q}) +
	h_{10}(\alpha_g(1-\alpha_g) - \frac32 (\alpha_{\bar q}^2+ \alpha_q^2)) \right]
\nonumber \\
{\cal A}_\perp (\alpha_i) &=& 30 \alpha_g^2(\alpha_{\bar q} - \alpha_q) \left[ h_{00} + h_{01} \alpha_g + \frac12 h_{10}(5 \alpha_g-3) \right]
\nonumber \\
B(u)&=& g_{\cal M}(u) - \phi_{\cal M}(u)
\nonumber \\
g_{\cal M}(u) &=& g_0 C_0^{\frac12}(2 u - 1) + g_2 C_2^{\frac12}(2 u - 1) + g_4 C_4^{\frac12}(2 u - 1)
\nonumber \\
{\mathbb A}(u) &=& 6 u \bar u \left[\frac{16}{15} + \frac{24}{35} a_2^{\cal M}+ 20 \eta_3 + \frac{20}{9} \eta_4 +
	\left( - \frac{1}{15}+ \frac{1}{16}- \frac{7}{27}\eta_3 w_3 - \frac{10}{27} \eta_4 \right) C_2^{3 \over 2}(2 u - 1) 
	\right. \nonumber \\ 
	&+& \left. \left( - \frac{11}{210}a_2^{\cal M} - \frac{4}{135} \eta_3w_3 \right)C_4^{3 \over 2}(2 u - 1)\right]
\nonumber \\
&+& \left( -\frac{18}{5} a_2^{\cal M} + 21 \eta_4 w_4 \right)\left[ 2 u^3 (10 - 15 u + 6 u^2) \ln u 
\right. \nonumber \\
&+& \left. 2 \bar u^3 (10 - 15 \bar u + 6 \bar u ^2) \ln\bar u + u \bar u (2 + 13 u \bar u) \right] 
\label{wavefns}
\end{eqnarray}
where $C_n^k(x)$ are the Gegenbauer polynomials,  
\begin{eqnarray}
h_{00}&=& v_{00} = - \frac13\eta_4
\nonumber \\
a_{10} &=& \frac{21}{8} \eta_4 w_4 - \frac{9}{20} a_2^{\cal M}
\nonumber \\
v_{10} &=& \frac{21}{8} \eta_4 w_4
\nonumber \\
h_{01} &=& \frac74  \eta_4 w_4  - \frac{3}{20} a_2^{\cal M}
\nonumber \\
h_{10} &=& \frac74 \eta_4 w_4 + \frac{3}{20} a_2^{\cal M}
\nonumber \\
g_0 &=& 1
\nonumber \\
g_2 &=& 1 + \frac{18}{7} a_2^{\cal M} + 60 \eta_3  + \frac{20}{3} \eta_4
\nonumber \\
g_4 &=&  - \frac{9}{28} a_2^{\cal M} - 6 \eta_3 w_3
\label{param0}
\end{eqnarray}
and  
the parameters entering Eqs. (\ref{wavefns}) and (\ref{param0}) are given in Table (\ref{param}) for the pion and the kaon.
\begin{table}
\begin{center}
\begin{tabular}{|c|c|c|}
\hline
		&	$\pi$	&	K	\\
\hline
$a_1^{\cal M}$	&	$0$	&	$0.050$	\\
\hline
$a_2^{\cal M}$	&	$0.44$ 	&	$0.16$	\\
\hline
$\eta_3$	&	$0.015$	&	$0.015$	\\
\hline
$\eta_4$	&	$10$	&	$0.6$	\\
\hline
$w_3$		&	$-3$	&	$-3$	\\
\hline
$w_4$		&	$0.2$	&	$0.2$	\\
\hline
\end{tabular}
\end{center}
\caption{Parameters of the wave function calculated at the renormalization scale $\mu = 1 ~GeV^2$}
\label{param}
\end{table}

Since the mass of the initial and final baryons are close to each other, so we can set $M_1^2 = M_2^2 = 2 M^2$. Then 
$u_0 = \frac12$. For this reason we need the value of the wave functions only at $u=\frac12$. The values of the other input
parameters appearing in the sum rules are: $\langle \bar q q \rangle = - (0.243~GeV)^3$, 
$m_0^2=(0.8 \pm 0.2)~GeV^2$ \cite{R23}, $f_K=0.160~GeV$, $f_\pi=0.131~GeV$ \cite{R21}

An upper bound for the Borel parameter $M^2$ is obtained by  requiring that the contribution of the continuum to the
correlation function is less then $50\%$ of the value of the correlation function. A lower bound is obtained by
requiring that the contribution of the term with the highest power of $\frac{1}{M^2}$  is less then $20\%$. Using these constraints,
one can find a working region for the Borel parameter $M^2$. Continuum threshold is varied in the range
between $s_0 = (m_B + 0.5)^2$ and $s_0  = (m_B + 0.7)^2$.

In Fig. (\ref{lambdaneutronK.Msq}), we present the dependence of $g_{\Lambda n K}$ on $M^2$ at three fixed values of the
parameter $t$ and two fixed values of the continuum threshold, $s_0$. From this figure, we see that the results are rather stable with respect to 
variations of $M^2$ in the working region of $M^2$. It is also seen that the result depends on the value of $t$. Our next task is to find
a region of $t$ where the results are independent of the value of $t$. For this aim, in Fig. (\ref{lambdaneutronK.th}), we depict the dependence
of $g_{\Lambda n K}$ on $\cos \theta$, where $\theta$ is defined through $t=\tan \theta$. We see that when $\cos \theta$ varies in between $-0.5 < \cos\theta < 0.25,$
the coupling constant $g_{\Lambda n K}$ is practically independent of the unphysical parameter $t$. And we find that $g_{\Lambda n K} = - 13\pm3$.

A similar analysis of the stability of the other coupling constants with respect to the variation of the Borel mass $M^2$, the parameter $t$ and the continuum 
threshold $s_0$ is performed (see Figs. (\ref{lambdasigmappi.Msq}-\ref{xi0xi0pi.th})). The results on the kaon and pion couplings are presented in Table \ref{results}
under the column ``General Current''. In the second column, labeled ``Ioffe current,'' are listed our predictions if we set the arbitrary parameter
$t=-1$. In the third column labeled ``$SU(3)_f,$'' we present the predictions of $SU(3)_f$ if one uses the central values of the prediction of the general current
for the transitions $p \rightarrow p \pi^0$ and $\Sigma^+ \rightarrow \Lambda \pi^+$ (marked as ``Input'' in the table) to determine
the $F$ and $D$ values.
In the same table, we also present the existing theoretical and experimental
results in the literature (columns $5$, $6$ and $7$)
for the same coupling constants.

Comparing the results in Table \ref{results}, we obtain the following main conclusions:
\begin{itemize}
\item
There are many cases in which the predictions of our analysis using the most general current differs considerable
from the prediction of the Ioffe current, e.g for the $\Lambda \rightarrow \Xi^0 K$ and $\Xi^0 \rightarrow \Lambda K^0$
coupling constants, the magnitudes as well as the signs are different. Also for the $p \rightarrow \Sigma^+ K^0$, $\Sigma^- \rightarrow n K^-$ and $n \rightarrow \Sigma^0
K^0$ cases, the magnitudes differ by at least a factor of three. This difference is mainly due to the fact that, in these decays, the predictions for the
coupling constants depend strongly on the exact value of $t$ around $t=-1$. Hence $t=-1$ does not fall in the stability region of the sum rules.

\item
Our results are also different from the results obtained in \cite{R14,R27} and \cite{R27p} 
but are closer to the experimental results \cite{R28} and \cite{R29}.

\item
In all cases except the $\Xi^0 \rightarrow \Xi^0 \pi^0$ transition,  the prediction of the general current is consistent with the $SU(3)_f$ symmetry, whereas
 the Ioffe current predicts large violation of $SU(3)_f$ symmetry. The connection between the $SU(3)_f$ symmetry and the usability of
 Ioffe current and the reason why there is a large violation of $SU(3)_f$ flavor symmetry in $\Xi^0 \rightarrow \Xi^0 \pi^0$ transition needs
 further study and is beyond the scope of this work. A plausibility argument can be that since $\Xi$ baryons contain two strange quarks, it is
 reasonable to expect that the $SU(3)_f$ violation will be more pronounced in this channel.
\end{itemize}

In conclusion, the coupling constants of the pseudo scalar $K$ and $\pi$ with 
the octet baryons are studied within the framework of
light cone QCD sum rules. In numerical analysis, we studied the $SU(3)_f$ flavor symmetry breaking effects due to 
$m_s \neq m_u=m_d=0$ and $\sp \neq \uu = \dd$. It is found that, without assuming any symmetry between the light quarks, all the
coupling constant of the octet baryons to $K$ and $\pi$ can be written in terms of only four analytical functions, which reduces
to the well known $F$ and $D$ functions in the $SU(3)_f$ symmetry case. We also perform comparison of our results with the existing theoretical and experimental
results in the literature.

\section*{Acknowledgments}
The work of T.M.A  is partially supported by TUBITAK under the project 105T131, and A.O. would like to thank TUBA-GEBIP for their partial financial support.

\begin{table}
\begin{center}
\small
\hspace*{-3.5cm}
\begin{tabular}{|c|c|c|c|c|c|c|}
\hline
Channel 			& General Current & Ioffe Current($t=-1$) & $SU(3)_f$ & QSR$^*$ & QSR$^\dagger$\cite{R27} & Exp.\\
\hline
$\Lambda \rightarrow n K$ 		& $-13\pm3$	& $-9.5\pm1$	& $-14.3$ &	$-2.37\pm0.09$\cite{R14}	&	$-2.49\pm1.25$	&	$-13.5$	\cite{R28}	\\
\hline
$\Lambda \rightarrow \Sigma^+ \pi^-$ 	& $10\pm3$	& $12\pm1$	& $10.0$ &	&	&	\\
\hline
$\Lambda \rightarrow \Xi^0 K^0$		& $4.5\pm2$	& $-2.5\pm0.5$	& $4.25$ &	&	&	\\
\hline
$n \rightarrow p \pi^-$			& $21\pm4$	& $20 \pm 2$	& $19.8$ &	&	& $21.2$ \cite{R29}	\\
\hline
$n \rightarrow \Sigma^0 K^0$		& $-3.2\pm2.2$	& $-9.5 \pm 0.5$& $- 3.3$ & $-0.025\pm0.015$\cite{R14}	& $-0.40 \pm 0.38$	& $-4.25$ \cite{R28}	\\
\hline
$p \rightarrow \Lambda K^+$		& $-13\pm3$	& $-10\pm1$	& $- 14.25$ & $-2.37\pm0.09$\cite{R14}	& $-2.49\pm1.25$	&$-13.5$ \cite{R28}	\\
\hline
$p \rightarrow p \pi^0$			& $14\pm4$	& $15 \pm 1$	& Input&$13.5\pm0.5$ \cite{R10}	&	& $14.9$ \cite{R29}	\\
\hline
$p \rightarrow \Sigma^+ K^0$ 		& $4\pm3$	& $14\pm1$	& $5.75$ &	&	&	\\
\hline
$\Sigma^0 \rightarrow n K^0$		& $-4 \pm 3$	& $-9.5\pm1$	&$-3.32$&	$-0.025\pm0.015$\cite{R14}	&	$-0.40\pm0.38$	&	$-4.25$	\cite{R28} \\
\hline
$\Sigma^0 \rightarrow \Lambda \pi^0$	&$11\pm3$ &	$12\pm1.5$ & $10.0$ & $6.9 \pm 1$ \cite{R27p}
& & \\
\hline
$\Sigma^0 \rightarrow \Xi^0 K^0$	& $-13\pm3$	& $-13.5\pm1$	& $-14$ & & &\\
\hline
$\Sigma^- \rightarrow n K^-$		& $5\pm3$	& $15\pm2$	& $4.7$ & & &\\
\hline
$\Sigma^+ \rightarrow	\Lambda \pi^+$	& $10\pm3.5$	& $12.5\pm1$	& Input & & &\\
\hline
$\Sigma^+ \rightarrow \Sigma^0 \pi^+$	& $-9\pm2$	& $-7.5\pm0.7$	& $-10.7$ &$-11.9\pm0.4$\cite{R10} & &\\
\hline
$\Xi^0 \rightarrow \Lambda K^0$		& $4.5\pm1$	& $-2.6\pm0.3$	& $4.25$ & & & \\
\hline
$\Xi^0 \rightarrow \Sigma^0 K^0$	& $-12.5\pm3$	& $-13.5\pm1$	& $-14$ & & &\\
\hline
$\Xi^0	\rightarrow \Sigma^+ K^-$	& $18\pm4$	& $19\pm2$	& $19.8$ & & &\\
\hline
$\Xi^0 \rightarrow \Xi^0 \pi^0$		& $10\pm2$	& $0.3\pm0.6$	& $-3.32$ & $-1.60\pm0.05$ \cite{R10}& &\\
\hline
\end{tabular}
\end{center}
\caption{The strong coupling constants for various channels both for the general current and for the Ioffe current. The first three columns are the results of
this work. QSR$^*$($^\dagger$) is the predictions of the QCD Sum Rules using the $\sigma_{\mu \nu} \gamma_5 p^\mu q^\nu$ ($i
\not\!q \gamma_5$) structure.}
\label{results}
\end{table}
\newpage
\begin{appendix}
\section{Expressions of the Correlation Functions}
Correlation functions for the couplings involving $\pi^0$:
\begin{eqnarray}
\Pi^{\Sigma^0 \rightarrow \Sigma^0 \pi^0} &=& \frac{1}{\sqrt2} \left( \Pi_1(u,d,s) - \Pi_1(d,u,s) \right)
\nonumber \\
\Pi^{\Sigma^+ \rightarrow \Sigma^+ \pi^0} &=& \sqrt2 \Pi_1(u,u,s)
\nonumber \\
\Pi^{\Sigma^- \rightarrow \Sigma^- \pi^0} &=& - \sqrt2 \Pi_1(d,d,s)
\nonumber \\
\Pi^{p \rightarrow p \pi^0} &=& \sqrt2 \Pi_1(u,u,d) - \frac1{\sqrt2} \Pi_2(u,u,d)
\nonumber \\
\Pi^{n \rightarrow n \pi^0} &=& - \sqrt2 \Pi_1(d,d,u) + \frac1{\sqrt2} \Pi_2(d,d,u)
\nonumber \\
\Pi^{\Xi^0 \rightarrow \Xi^0 \pi^0} &=& \frac1{\sqrt2} \Pi_2(s,s,u)
\nonumber \\
\Pi^{\Xi^- \rightarrow \Xi^- \pi^0} &=&- \frac1{\sqrt2} \Pi_2(s,s,d)
\nonumber \\
\Pi^{\Sigma^0 \rightarrow \Lambda \pi^0} + \Pi^{\Lambda \rightarrow \Sigma^0 \pi^0} &=&
\frac2{\sqrt6} \left[ \Pi_1(u,s,d) + \Pi_1(d,s,u) - \Pi_2(s,d,u) - \Pi_2(s,u,d)\right]
\nonumber \\
\Pi^{\Lambda \rightarrow \Lambda \pi^0} &=& 
\frac{\sqrt2}{3} \left[ \Pi_1(u,s,d) - \Pi_1(d,s,u) + \Pi_2(s,d,u) 
\right. \nonumber \\ 
&& - \left. \Pi_2(s,u,d)
- \frac12 \Pi_1(u,d,s) + \frac12 \Pi_1(d,u,s) \right]
\end{eqnarray} 
Correlation functions for the couplings involving $\pi^-$:
\begin{eqnarray}
\Pi^{\Sigma^0 \rightarrow \Sigma^+ \pi^-} &=& -\sqrt2 \Pi_1(d,u,s)
\nonumber \\
\Pi^{\Sigma^- \rightarrow \Sigma^0 \pi^-} &=& \sqrt2 \Pi_1(u,d,s)
\nonumber \\
\Pi^{\Xi^- \rightarrow \Xi^0 \pi^-} &=& - \Pi_2(s,s,d)
\nonumber \\
\Pi^{\Lambda \rightarrow \Sigma^+ \pi^-} &=&-\frac1{\sqrt3} \left[ 2 \Pi_3(d,s,u) + \sqrt2 \Pi_1(d,u,s) \right]
\nonumber \\
\Pi^{\Sigma^- \rightarrow \Lambda \pi^-} &=& - \frac1{\sqrt3} \left[ 2 \Pi_4(u,s,d) + \sqrt2 \Pi_1(u,d,s) \right]
\nonumber \\
\Pi^{n \rightarrow p \pi^-} &=&  -\sqrt2\Pi_3(d,d,u)
\end{eqnarray}
Correlation functions for the couplings involving $\pi^+$:
\begin{eqnarray}
\Pi^{\Sigma^+  \rightarrow \Sigma^0 \pi^+} &=& -\sqrt2 \Pi_1(d,u,s)
\nonumber \\
\Pi^{\Sigma^0 \rightarrow \Sigma^- \pi^+} &=& \sqrt2 \Pi_1(u,d,s)
\nonumber \\
\Pi^{\Xi^0 \rightarrow \Xi^- \pi^+} &=& - \Pi_2(s,s,u)
\nonumber \\
\Pi^{\Sigma^+ \rightarrow \Lambda \pi^+} &=&  -\frac1{\sqrt3} \left[ 2 \Pi_4(d,s,u) + \sqrt2 \Pi_1(d,u,s) \right]
\nonumber \\
\Pi^{\Lambda \rightarrow \Sigma^- \pi^+} &=& - \frac1{\sqrt3} \left[ 2 \Pi_3(u,s,d) + \sqrt2 \Pi_1(u,d,s) \right]
\nonumber \\
\Pi^{p \rightarrow n \pi^+} &=& -\sqrt2 \Pi_3(u,u,d)
\end{eqnarray}
Correlation functions for the couplings involving $K^+$
\begin{eqnarray}
\Pi^{p \rightarrow \Sigma^0 K} &=& -\sqrt2 \Pi_1(s,u,d) - \Pi_4(s,d,u)
\nonumber \\
\Pi^{p \rightarrow \Lambda K} &=& -\frac1{\sqrt3} \left[ \sqrt2 \Pi_1(s,u,d) - \Pi_4(s,d,u) \right]
\nonumber \\
\Pi^{n \rightarrow \Sigma^- K} &=& - \Pi_2(d,d,u)
\nonumber \\
\Pi^{\Sigma^+ \rightarrow \Xi^0 K} &=& -\sqrt2\Pi_3(u,u,s)
\nonumber \\
\Pi^{\Sigma^0 \rightarrow \Xi^- K} &=& -\Pi_3(u,d,s)
\nonumber \\
\Pi^{\Lambda \rightarrow  \Xi^- K} &=&  \frac1{\sqrt3} \left[ 2 \sqrt2 \Pi_1(u,s,d) + \Pi_3(u,d,s) \right]
\end{eqnarray}
Correlation functions for the couplings involving $K^-$
\begin{eqnarray}
\Pi^{\Sigma^0  \rightarrow p K} &=& -\sqrt2 \Pi_1(s,u,d) - \Pi_3(s,d,u)
\nonumber \\
\Pi^{\Lambda \rightarrow p K} &=&- \frac1{\sqrt3}\left[ \sqrt2 \Pi_1(s,u,d) - \Pi_3(s,d,u) \right]
\nonumber \\
\Pi^{\Sigma^- \rightarrow n K} &=& - \Pi_2(d,d,s)
\nonumber \\
\Pi^{\Xi^0 \rightarrow \Sigma^+ K} &=& -\sqrt2 \Pi_3(s,s,u)
\nonumber \\
\Pi^{\Xi^- \rightarrow \Sigma^0 K} &=& -\Pi_4(u,d,s)
\nonumber \\
\Pi^{\Xi^- \rightarrow \Lambda K} &=&  \frac1{\sqrt3} \left[ 2 \sqrt2 \Pi_1(u,s,d) + \Pi_4(u,d,s) \right]
\end{eqnarray} 
Correlation functions for the couplings involving $K^0(s \bar d)$
\begin{eqnarray}
\Pi^{\Xi^0 \rightarrow \Sigma^0 K} &=& \Pi_4(d,u,s)
\nonumber \\
\Pi^{\Xi^0 \rightarrow \Lambda K} &=& \frac1{\sqrt3} \left[2 \sqrt2 \Pi_1(d,s,u) + \Pi_4(d,u,s) \right]
\nonumber \\
\Pi^{\Xi^- \rightarrow \Sigma^- K} &=& -\sqrt2 \Pi_3(s,s,d)
\nonumber \\
\Pi^{\Sigma^0 \rightarrow n K} &=& \Pi_3(s,u,d) + \sqrt2 \Pi_1(s,d,u)
\nonumber \\
\Pi^{\Lambda \rightarrow n K} &=& \frac1{\sqrt3} \left[ \Pi_3(s,u,d) - \sqrt2 \Pi_1(s,d,u) \right]
\nonumber \\
\Pi^{\Sigma^+ \rightarrow p K} &=& - \Pi_2(u,u,s)
\end{eqnarray} 
Correlation functions for the couplings involving $\bar{K^0}(d \bar s)$
\begin{eqnarray}
\Pi^{\Sigma^0 \rightarrow \Xi^0 K} &=& \Pi_3(d,u,s)
\nonumber \\
\Pi^{\Lambda \rightarrow \Xi^0 K} &=& \frac1{\sqrt3} \left[ 2 \sqrt2 \Pi_1(d,s,u) + \Pi_3(d,u,s) \right]
\nonumber \\
\Pi^{\Sigma^- \rightarrow \Xi^- K} &=& -\sqrt2 \Pi_3(d,d,s)
\nonumber \\
\Pi^{n \rightarrow \Sigma^0 K} &=&  \Pi_4(s,u,d) + \sqrt2 \Pi_1(s,d,u) 
\nonumber \\
\Pi^{n \rightarrow \Lambda K} &=& \frac1{\sqrt3} \left[ \Pi_4(s,u,d) - \sqrt2 \Pi_1(s,d,u) \right]
\nonumber \\
\Pi^{p \rightarrow \Sigma^+ K} &=&- \Pi_2(u,u,d)
\end{eqnarray} 
\section{Relations in the $SU(2)$ Limit}
Correlation functions involving the pion:
\begin{eqnarray}
\Pi^{\Sigma^0 \rightarrow \Sigma^0 \pi} &=& \Pi^{\Lambda \rightarrow \Lambda \pi^0} = 0
\nonumber \\
\sqrt2 \Pi_1(q,q,s) &=& \Pi^{\Sigma^+ \rightarrow \Sigma^+ \pi} = - \Pi^{\Sigma^- \rightarrow \Sigma^- \pi} = - \Pi^{\Sigma^0 \rightarrow 
\Sigma^+ \pi} 
\nonumber \\
&=& \Pi^{\Sigma^- \rightarrow \Sigma^0 \pi} = - \Pi^{\Sigma^+ \rightarrow \Sigma^0 \pi} = \Pi^{\Sigma^0 \rightarrow \Sigma^- \pi}
\nonumber \\
\Pi^{\Xi^0 \rightarrow \Xi^0 \pi} &=& \frac1{\sqrt2} \Pi_2(s,s,q) = - \Pi^{\Xi^- \rightarrow \Xi^- \pi}= - \frac{1}{\sqrt2} \Pi^{\Xi^- \rightarrow \Xi^0 \pi}
= - \frac{1}{\sqrt2} \Pi^{\Xi^0 \rightarrow \Xi^- \pi}
\nonumber \\
\Pi^{p \rightarrow p \pi} &=& - \Pi^{n \rightarrow n \pi} = \sqrt2 \Pi_1(q,q,q) - \frac1{\sqrt2} \Pi_2(q,q,q)
\nonumber  \\
\Pi^{\Lambda \rightarrow \Sigma^+ \pi} &=& \Pi^{\Lambda \rightarrow \Sigma^- \pi} = -\frac1{\sqrt3} \left[ 2 \Pi_3(q,s,q) + \sqrt2 \Pi_1(q,q,s) \right]
\nonumber \\
\Pi^{\Sigma^+ \rightarrow \Lambda \pi} &=& \Pi^{\Sigma^- \rightarrow \Lambda \pi} =- \frac1{\sqrt3} \left[ 2 \Pi_4(q,s,q) + \sqrt2 \Pi_1(q,q,s) \right]
\nonumber \\
\Pi^{n \rightarrow p \pi} &=& \Pi^{p \rightarrow n \pi} = -\sqrt2\Pi_3(q,q,q)
\nonumber \\
\Pi^{\Sigma^0  \rightarrow \Lambda \pi} + \Pi^{\Lambda \rightarrow \Sigma^0 \pi^0} &=& \frac{4}{\sqrt6} \left[ \Pi_1(q,s,q) - \Pi_2(s,q,q) \right]
\end{eqnarray}
Correlations involving the kaons:
\begin{eqnarray}
\Pi^{n \rightarrow \Sigma^0 K} &=& - \Pi^{p \rightarrow \Sigma^0 K} = \Pi_4(s,q,q) + \sqrt2 \Pi_1(s,q,q)
\nonumber \\ 
\Pi^{p \rightarrow \Lambda K} &=& \Pi^{n \rightarrow \Lambda K} = -\frac1{\sqrt3} \left[ \sqrt2 \Pi_1(s,q,q) - \Pi_4(s,q,q) \right]
\nonumber \\
\Pi^{p \rightarrow \Sigma^+K} &=& \Pi^{n \rightarrow \Sigma^- K} = - \Pi_2(q,q,q)
\nonumber \\
-\Pi^{\Sigma^0 \rightarrow \Xi^- K} &=& -\frac{1}{\sqrt2} \Pi^{\Sigma^+ \rightarrow \Xi^0 K} = \Pi^{\Sigma^0 \rightarrow \Xi^0 K} = -\frac{1}{\sqrt2} \Pi^{\Sigma^- \rightarrow
\Xi^- K} = \Pi_3(q,q,s)
\nonumber \\
\Pi^{\Lambda \rightarrow \Xi^0 K} &=&  \Pi^{\Lambda \rightarrow \Xi^- K} = \frac1{\sqrt3} \left[ 2 \sqrt2 \Pi_1(q,s,q) + \Pi_3(q,q,s) \right]
\nonumber \\
\Pi^{\Sigma^0 \rightarrow n K} &=& -\Pi^{\Sigma^0 \rightarrow p K} = \sqrt2 \Pi_1(s,q,q) + \Pi_3(s,q,q) 
\nonumber \\
-\Pi^{\Lambda \rightarrow p K} &=& - \Pi^{\Lambda \rightarrow n K} = \frac1{\sqrt3}\left[ \sqrt2 \Pi_1(s,q,q) - \Pi_3(s,q,q) \right]
\nonumber \\
\Pi^{\Sigma^- \rightarrow n K} &=& \Pi^{\Sigma^+ \rightarrow p K} = - \Pi_2(q,q,s)
\nonumber \\
-\Pi^{\Xi^0 \rightarrow \Sigma^+ K} &=&- \Pi^{\Xi^- \rightarrow \Sigma^- K} = \sqrt2 \Pi_3(s,s,q)
\nonumber \\ 
-\Pi^{\Xi^- \rightarrow \Sigma^0 K} &=& \Pi^{\Xi^0 \rightarrow \Sigma^0 K} = \Pi_4(q,q,s)
\nonumber \\
\Pi^{\Xi^- \rightarrow \Lambda K} &=& \Pi^{\Xi^0 \rightarrow \Lambda K} = \frac1{\sqrt3} \left[ 2 \sqrt2 \Pi_1(q,s,q) + \Pi_4(q,q,s) \right]
\end{eqnarray}

\end{appendix}

\newpage

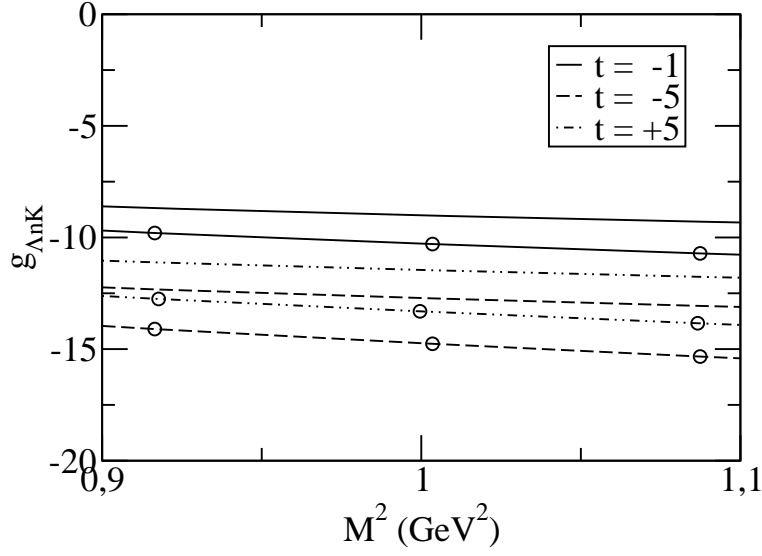
\figt{lambda.neutron.K.Msq.eps}{lambdaneutronK.Msq}{The dependence of the
coupling constant $g_{\Lambda n K}$ on the Borel parameter $M^2$ for
the values of the arbitrary parameter $t = -1, \pm5$ and the continuum
threshold $s_0=2.25~GeV^2$(the curves without any symbols) and 
$s_0=2.75~GeV^2$(the curves with circles).}

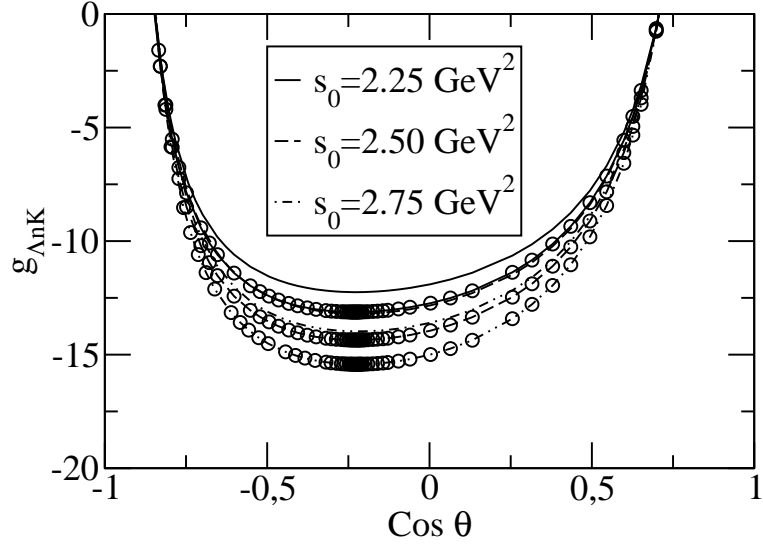
\figb{lambda.neutron.K.th.eps}{lambdaneutronK.th}{The dependence of the
coupling constant $g_{\Lambda n K}$ on $\cos \theta$ for $s_0= 2.50 \pm
0.25~GeV^2$ and for the Borel parameter $M^2=0.9~GeV^2$(the curves without
ant symbols) and $M^2=1.1~GeV^2$(the curves with circles on them).}

\newpage 

\figt{lambda.sigmap.pi.Msq.eps}{lambdasigmappi.Msq}{Same as Fig. \ref{lambdaneutronK.Msq} but for the $\Lambda \rightarrow 
\Sigma^+ \pi^-$ transition}

\figb{lambda.sigmap.pi.th.eps}{lambdasigmappi.th}{Same as Fig. \ref{lambdaneutronK.th} but for the $\Lambda \rightarrow 
\Sigma^+ \pi^-$ transition}

\newpage
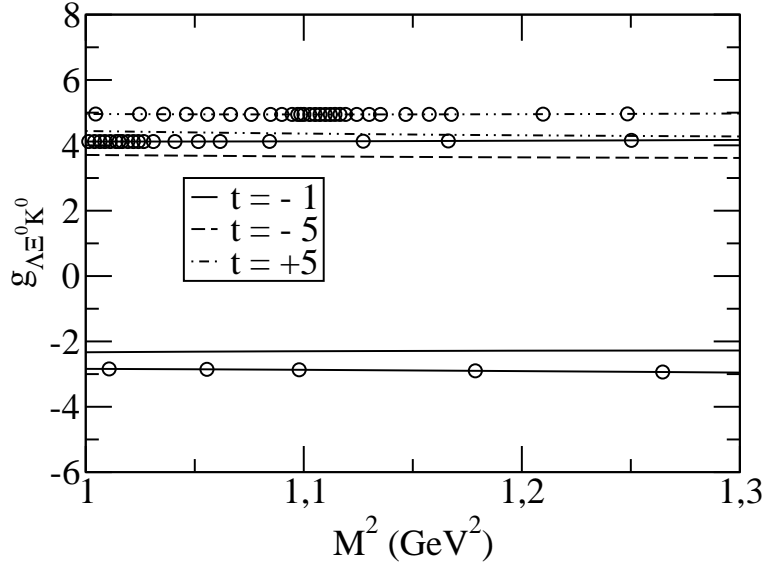
\figt{lambda.xi0.K.Msq.eps}{lambdaxi0K.Msq}{Same as Fig. \ref{lambdaneutronK.Msq} but for the 
$\Lambda \rightarrow \Xi^0 K^0$ transition and for the threshold values $s_0=2.75$(the curves without any symbols) and
$s_0=3.25$ (curves with circles)}

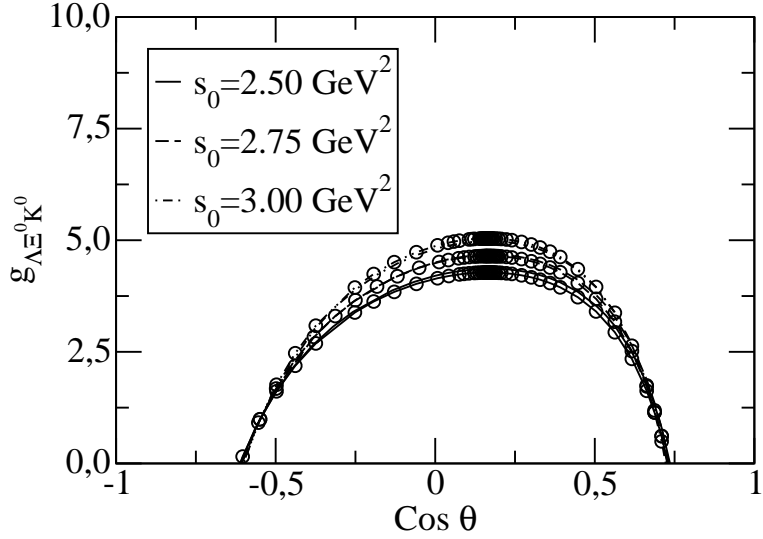
\figb{lambda.xi0.K.th.eps}{lambdaxi0K.th}{Same as Fig. \ref{lambdaneutronK.th} but for the 
$\Lambda \rightarrow \Xi^0 K^0$ transition and for the values of the threshold $s_0=2.75\pm0.25~GeV^2$ and the Borel mass
$M^2=1.1$(curves  without any circles), and
$M^2=1.3$(curves with circles)}

\newpage
\figt{neutron.proton.pi.Msq.eps}{neutronprotonpi.Msq}{The same as Fig. \ref{lambdaneutronK.Msq} but for the 
$n \rightarrow p \pi^-$ transition.}

\figb{neutron.proton.pi.th.eps}{neutronprotonpi.th}{The same as Fig. \ref{lambdaneutronK.th} but for the
$n \rightarrow p \pi^-$ transition.}

\newpage
\figt{neutron.sigma0.K.Msq.eps}{neutronsigma0K.Msq}{The same as Fig. \ref{lambdaneutronK.Msq} but for the 
$n \rightarrow \Sigma^0 K^0$ transition.}

\figb{neutron.sigma0.K.th.eps}{neutronsigma0K.th}{The same as Fig. \ref{lambdaneutronK.th} but for the 
$n \rightarrow \Sigma^0 K^0$ transition.}

\newpage
\figt{proton.lambda.K.Msq.eps}{protonlambdaK.Msq}{The same as Fig. \ref{lambdaneutronK.Msq} but for the 
$p \rightarrow \Lambda K^+$ transition.}

\figb{proton.lambda.K.th.eps}{protonlambdaK.th}{The same as Fig. \ref{lambdaneutronK.th} but for the 
$p \rightarrow \Lambda K^+$ transition.}

\newpage

\figt{proton.proton.pi.Msq.eps}{protonprotonpi.Msq}{The same as Fig. \ref{lambdaneutronK.Msq} but for the 
$p \rightarrow p \pi^0$ transition.}

\figb{proton.proton.pi.th.eps}{protonprotonpi.th}{The same as Fig. \ref{lambdaneutronK.th} but for the 
$p \rightarrow p \pi^0$ transition.}

\newpage
\figt{proton.sigmap.K.Msq.eps}{protonsigmapK.Msq}{The same as Fig. \ref{lambdaneutronK.Msq} but for the 
$p \rightarrow \Sigma^+ K^0$ transition.}

\figb{proton.sigmap.K.th.eps}{protonsigmapK.th}{The same as Fig. \ref{lambdaneutronK.th} but for the 
$p \rightarrow \Sigma^+ K^0$ transition.}

\newpage
\figt{sigma0.neutron.K.Msq.eps}{sigma0neutronK.Msq}{The same as Fig. \ref{lambdaneutronK.Msq} but for the 
$\Sigma^0 \rightarrow n K^0$ transition.}

\figb{sigma0.neutron.K.th.eps}{sigma0neutronK.th}{The same as Fig. \ref{lambdaneutronK.th} but for the 
$\Sigma^0 \rightarrow n K^0$ transition.}

\clearpage
\newpage
\figt{sigma0.lambda.pi.Msq.eps}{sigma0lambdapi.Msq}{The same as Fig. \ref{lambdaneutronK.Msq} but for the 
$\Sigma^0 \rightarrow \Lambda \pi^0$ transition.}

\figb{sigma0.lambda.pi.th.eps}{sigma0lambdapi.th}{The same as Fig. \ref{lambdaneutronK.th} but for the 
$\Sigma^0 \rightarrow \Lambda \pi^0$ transition.}

\clearpage
\newpage
\figt{sigma0.xi0.K.Msq.eps}{sigma0xi0K.Msq}{The same as Fig. \ref{lambdaxi0K.Msq} but for the 
$\Sigma^0 \rightarrow \Xi^0 K^0$ transition.}

\figb{sigma0.xi0.K.th.eps}{sigma0xi0K.th}{The same as Fig. \ref{lambdaxi0K.th} but for the 
$\Sigma^0 \rightarrow \Xi^0 K^0$ transition.}

\newpage

\figt{sigmam.neutron.K.Msq.eps}{sigmamneutronK.Msq}{The same as Fig. \ref{lambdaxi0K.Msq} but for the 
$\Sigma^- \rightarrow n K^-$ transition.}

\figb{sigmam.neutron.K.th.eps}{sigmamneutronK.th}{The same as Fig. \ref{lambdaxi0K.th} but for the 
$\Sigma^- \rightarrow n K^-$ transition.}

\newpage
\figt{sigmap.lambda.pi.Msq.eps}{sigmaplambdapi.Msq}{The same as Fig. \ref{lambdaneutronK.Msq} but for the 
$\Sigma^+ \rightarrow \Lambda  \pi^+$ transition.}

\figb{sigmap.lambda.pi.th.eps}{sigmaplambdapi.th}{The same as Fig. \ref{lambdaneutronK.th} but for the 
$\Sigma^+ \rightarrow \Lambda  \pi^+$ transition.}

\newpage

\figt{sigmap.sigma0.pi.Msq.eps}{sigmapsigma0pi.Msq}{The same as Fig. \ref{lambdaneutronK.Msq} but for the 
$\Sigma^+ \rightarrow \Sigma^0 \pi^+$ transition.}

\figb{sigmap.sigma0.pi.th.eps}{sigmapsigma0pi.th}{The same as Fig. \ref{lambdaneutronK.th} but for the 
$\Sigma^+ \rightarrow \Sigma^0 \pi^+$ transition.}

\newpage

\figt{xi0.lambda.K.Msq.eps}{xi0lambdaK.Msq}{The same as Fig. \ref{lambdaxi0K.Msq} but for the 
$\Xi^0 \rightarrow \Lambda  K^0$ transition.}

\figb{xi0.lambda.K.th.eps}{xi0lambdaK.th}{The same as Fig. \ref{lambdaxi0K.th} but for the 
$\Xi^0 \rightarrow \Lambda  K^0$ transition.}

\newpage

\figt{xi0.sigma0.K.Msq.eps}{xi0sigma0K.Msq}{The same as Fig. \ref{lambdaxi0K.Msq} but for the 
$\Xi^0 \rightarrow \Sigma^0 K^0$ transition.}

\figb{xi0.sigma0.K.th.eps}{xi0sigma0K.th}{The same as Fig. \ref{lambdaxi0K.th} but for the 
$\Xi^0 \rightarrow \Sigma^0 K^0$ transition.}

\newpage

\figt{xi0.sigmap.K.Msq.eps}{xi0sigmapK.Msq}{The same as Fig. \ref{lambdaxi0K.Msq} but for the 
$\Xi^0 \rightarrow \Sigma^+ K^-$ transition.}

\figb{xi0.sigmap.K.th.eps}{xi0sigmapK.th}{The same as Fig. \ref{lambdaxi0K.th} but for the 
$\Xi^0 \rightarrow \Sigma^+ K^-$ transition.}
\newpage

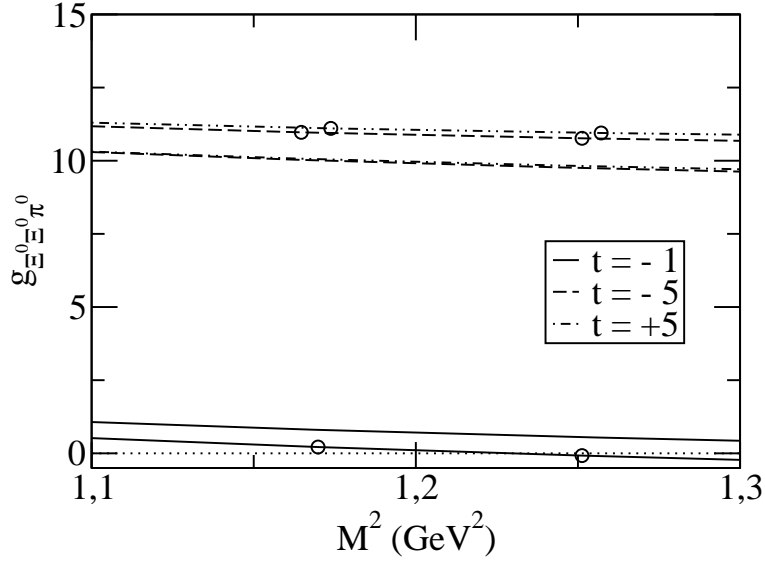
\figt{xi0.xi0.pi.Msq.eps}{xi0xi0pi.Msq}{The same as Fig. \ref{lambdaneutronK.Msq} but for the 
$\Xi^0 \rightarrow \Xi^0 \pi^0$ transition and the threshold values $s_0=2.75$(curves without any symbols), and
$s_0=3.25$(curves
with symbols).}

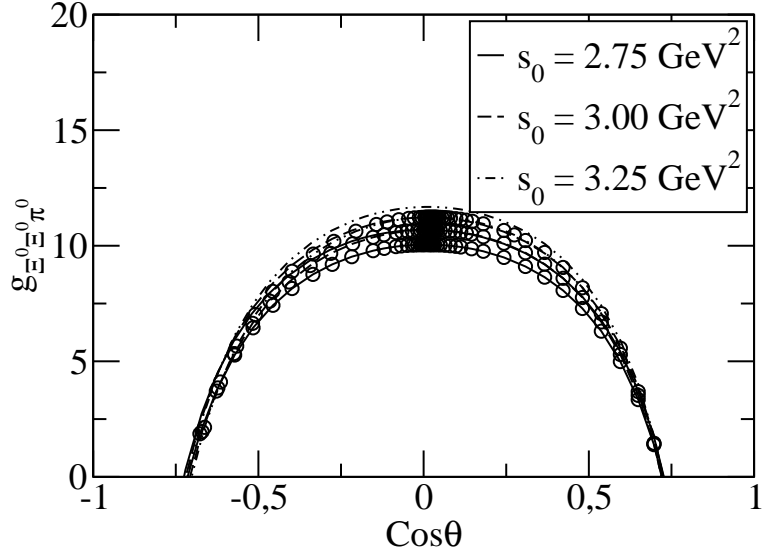
\figb{xi0.xi0.pi.th.eps}{xi0xi0pi.th}{Same as Fig. \ref{lambdaneutronK.th} but for the 
$\Xi^0 \rightarrow \Xi^0 \pi^0$ transition and for the values of the threshold $s_0=3.00\pm0.25~GeV^2$ and the Borel mass
$M^2=1.1$(curves  without any circles), and
$M^2=1.3$(curves with circles)}
\end{document}